\documentclass[fleqn,usenatbib]{mnras}

\usepackage{newtxtext,newtxmath}

\usepackage[T1]{fontenc}
\usepackage{CJKutf8}
\usepackage{orcidlink}

\DeclareRobustCommand{\VAN}[3]{#2}
\let\VANthebibliography\thebibliography
\def\thebibliography{\DeclareRobustCommand{\VAN}[3]{##3}\VANthebibliography}

\newcommand{\dphi}{$\mathrm{\Delta\phi}$}
\newcommand{\Myoung}{$M_{\mathrm{young}}$}
\newcommand{\Mstar}{$M_*$}
\newcommand{\Mgas}{$M_{\mathrm{gas}}$}
\newcommand{\Dsat}{$D_{\mathrm{sat}}$}
\newcommand{\Rvir}{$R_{\mathrm{200}}$}
\newcommand{\dage}{$\mathrm{\Delta\mathrm{age}}$}
\newcommand{\fno}{$f_{\mathrm{non-overlap}}$}
\newcommand{\dmu}{$\mathrm{\Delta\tau}$}
\newcommand{\cone}{case~\textsc{i}}
\newcommand{\ctwo}{case~\textsc{ii}}
\newcommand{\cthree}{case~\textsc{iii}}

\usepackage{graphicx}	
\usepackage{amsmath}	

\newcommand{\red}[1]{{\color{black}#1}}
\newcommand{\revi}[1]{{\color{black}#1}}

\title[Spiral galaxies in Auriga L3]{How Mergers and Flybys Shape Azimuthal Age Patterns in Spiral Galaxies}

\author[Qian-Hui Chen et al.]{\ignorespaces
Qian-Hui Chen (陈千惠)$^{\orcidlink{0000-0002-4382-1090}}$, $^{1,2}$\thanks{E-mail: Qianhui.Chen@anu.edu.au}
Alex M. Garcia$^{\orcidlink{0000-0002-8111-9884}}$, $^{3}$
Zefeng Li (李泽峰)$^{\orcidlink{0000-0001-7373-3115}}$, $^{4}$
Kathryn Grasha$^{\orcidlink{0000-0002-3247-5321}}$, $^{1,2}$\thanks{ARC DECRA Fellow}
\newauthor
Emily Wisnioski$^{\orcidlink{0000-0003-1657-7878}}$, $^{1,2}$
Paul Torrey$^{\orcidlink{0000-0002-5653-0786}}$, $^{3}$
Rhea-Silvia Remus$^{\orcidlink{0009-0008-9260-7278}}$, $^{5}$
Lucas C. Kimmig $^{\orcidlink{0009-0006-8337-8712}}$,$^{5}$
\newauthor
Andrew J. Battisti$^{\orcidlink{0000-0003-4569-2285}}$ $^{1,2,6}$
and
Sven Buder$^{\orcidlink{0000-0002-4031-8553}}$ $^{1,2}$\footnotemark[2]
\\
$^{1}$Research School of Astronomy and Astrophysics, Australian National University, Canberra, ACT 2611, Australia\\
$^{2}$ARC Centre of Excellence for All Sky Astrophysics in 3 Dimensions (ASTRO 3D), Australia\\
$^{3}$Department of Astronomy, University of Virginia, Charlottesville, VA 22904, USA\\
$^{4}$Centre for Extragalactic Astronomy, Department of Physics, Durham University, South Road, Durham DH1 3LE, UK\\
$^{5}$Universitäts-Sternwarte München, Fakultät für Physik, LMU München, Scheinerstr. 1, D-81679 München, Germany\\
$^{6}$International Centre for Radio Astronomy Research, University of Western Australia, 35 Stirling Highway, Crawley WA 6009, Australia\\
}

\date{Accepted XXX. Received YYY; in original form ZZZ}

\pubyear{\the\year{}}

\begin{document}
\begin{CJK}{UTF8}{gbsn}
	\label{firstpage}
	\pagerange{\pageref{firstpage}--\pageref{lastpage}}
	\maketitle

\begin{abstract}
Spiral structures are one of the most common features in galaxies, yet their origins and evolution remain debated. 
Stellar age distributions offer crucial insights into galaxy evolution and star formation, though environmental effects can obscure the intrinsic age patterns. 
Using the Auriga cosmological gravo-magnetohydrodynamical zoom-in simulations, we investigate the azimuthal age distribution of young stars ($<2$ Gyr) in a sample of five Milky Way-mass spiral galaxies over the past 5~Gyr. 
We quantify the age gradients across spiral arms using the mean age offset (\dmu) and the non-overlap fraction (\fno). 
We further analyse the impact of mergers and fly-by events on the age gradients. 
Our results show that Auriga spiral galaxies generally feature younger stars in their leading edges compared to the trailing edges, with a typical \dmu\ between 30 and 80~Myr. 
However, gas-rich interactions can disrupt this age offset, resulting in similar age distributions on each side of the spiral arms. 
In three snapshots, we observe similar mean ages on both sides of spiral arms but differing age distribution broadness, coinciding with satellite interactions crossing the host galaxy's disc plane. 
Our simulation data suggest that the typical azimuthal age variation recovers within $\sim$600~Myr after galaxy interactions.
This work highlights the transient role of environmental interactions in shaping spiral arm age patterns.

\end{abstract}

\begin{keywords}
galaxies: disc -- galaxies: spiral -- galaxies: stellar content -- galaxies: structure -- galaxies: evolution
\end{keywords}



\section{Introduction}\label{sec:intro}
Large image-based surveys \citep[e.g., Sloan Digital Sky Survey, hereafter SDSS;][]{York_2000} -- along with extensive analyses by collaborations such as Galaxy Zoo \citep{Lintott_2011} -- have revealed that approximately two-thirds of massive galaxies in the local Universe display spiral structures. 
Spiral galaxies typically exhibit higher star formation rates (SFRs) than elliptical galaxies, making them the primary hosts of star formation in the local Universe \citep{Martig_2013}. 
At higher redshifts (0.5 $\leq z \leq$ 4), \citet{Kuhn_2024} find that a quarter of galaxies (stellar masses \Mstar $> 10^{10}$ M$_\odot$) already display spiral-like structures.
Understanding the origin of spiral arms and their relations to star formation is thus crucial to unravelling how galaxies have evolved from high redshifts to their present-day forms.

Spiral arms can be classified into grand-design from a simple wave mode, and flocculent and multi-armed from various wave modes \citep{Elmegreen90, Elmegreen93}.
Disc instability is considered to be the major mechanism in the formation of grand design spirals \citep[e.g.][]{Toomre_1972, BT08}.
Within the disc instability paradigm, there are two main theories to explain the origin of spiral arms \citep{Dobbs_2014, Sellwood_2022} -- \textit{density wave theory} \citep{Lin_1966} and 
\textit{dynamic spiral theory} \citep{Goldreich_1965, Toomre_1977}.
The former does not necessarily depend on a globally unstable disc, while disc instability may amplify the effects; the latter is instead heavily reliant on local gravitational instabilities \citep[e.g.][]{Julian_1966}.
Disc instability is also associated with perturbation events such as a flyby interaction or minor merger (e.g. M51; see \citeauthor{BT08} \citeyear{BT08} and \citeauthor{Sellwood_2011} \citeyear{Sellwood_2011} for reviews).

The density wave theory proposes that long-lived spiral density waves propagate through a galaxy disc, rotating at a constant angular speed.
As molecular gas clouds enter density wave potentials, cloud-cloud collisions are triggered, which subsequently enhances SFRs \citep{Silva-Villa_2012, Cedres_2013, Choi_2015}. 
Since material in the disc rotates at decreasing angular velocities with increasing radii, newly formed stars move ahead of the density waves inside the co-rotation radius, where the disc materials and the spiral pattern rotate at the same rate. 
This differential rotation hypothesis predicts an age gradient across spiral arms, with young stars concentrated near the density waves and progressively older stars found further away \citep{Martinez-Garcia_2009, Sanchez-Gil_2011, Shabani_2018, Vallee_2022, Martinez-Garcia_2023}. 
Under this framework, the leading edge of a spiral arm is expected to host an intermediate-age stellar population, while the trailing edge of the following spiral pattern harbours the oldest stellar populations.

In contrast, the dynamic spiral theory offers a different perspective of stellar distributions in spiral galaxies \citep{Goldreich_1965, Julian_1966, Toomre_1977}.
According to this theory, spiral arms are short-lived and transient features that arise from local gravitational instabilities \citep{Marchuk_2018}. 
Galactic shear stretches and disrupts the spiral arms locally, while self-gravity works to hold the segments together.
This tug-of-war between shear and self-gravity causes the arms to fragment into structures that are a few kpc in size, which then reconnect to form larger-scale patterns \citep{Fujii_2011, Benhaiem_2019}.
Unlike the density wave theory, the dynamic spiral theory posits that gas flows into the potential wells of the spiral arms from both sides, resulting in no differential rotation between the spiral arms and the surrounding disc material.
Consequently, the dynamic spiral theory does not predict a systematic age gradient across the spiral arms.

Both the density wave theory and dynamic spiral theory effectively explain the formation of spiral structures in isolated galaxies, as supported by observations and simulations \citep{Dobbs_2010, Fujii_2011, Grand_2012, Baba_2013, Donghia_2013, Khrapov_2021, Kumar_2021}.
However, spiral structures are not exclusive to isolated galaxies and can also arise due to galaxy interactions, leading to the formation of tidally induced spiral arms. 
When a satellite galaxy passes by, tidal forces can create a bridge-and-tail structure \red{\citep{Toomre_1972}} that subsequently evolves into grand-design spiral patterns in the host galaxy. 
These tidal-driven spiral arms have been observed in the local Universe \citep{Darg_2010, Di_Teodoro_2014}, at $z\sim 0.3$ \citep[e.g., G12-J120759 in][]{Foster_2021} and near cosmic noon \citep[$1<z<3$;][]{Law_2012, Margalef-Bentabol_2022} in galaxies undergoing minor or major mergers.
Tidal-induced spiral arms can be distinguished from those formed by density wave or dynamic spiral theories through the presence of a companion galaxy \citep{Semczuk_2017, Antoja_2022}. 
Furthermore, bridge and tail structures associated with tidal interactions exhibit distinct kinematic properties, which differ from the predictions of both density wave and dynamic spiral theories \citep{Pettitt_2016, Pettitt_2017}.

Measuring azimuthal variations in stellar age and the interstellar medium (ISM) provides a straightforward approach to studying the influence of spiral arms on star formation propagation, though it comes with significant challenges. 
Some observational studies, e.g., \citet{Tamburro_2008} and \citet{Egusa_2009}, have reported spatial offsets between ionized and molecular gas, consistent with density wave theory. 
Similarly, azimuthal offsets are found in stellar age (\citeauthor{Sanchez-Gil_2011}, \citeyear{Sanchez-Gil_2011} and \citeauthor{Shabani_2018}\citeyear{Shabani_2018}).
In contrast, other studies \citep{Scheepmaker_2009, Kaleida_2010, Shabani_2018} report no systematic trend of age gradients across spiral arms, and \citet{Foyle_2011} found no consistent azimuthal offsets between different tracers in 12 nearby spiral galaxies. 
More recently, \citet{Chen_2024} examined two spiral galaxies at $z\sim0.3$ and found no offset in $\mathrm{D_{4000}}$, an age proxy measured from the spectral continuum, between the leading and trailing edges of the arms across various radii. 
\citet{Chen_2024} suggest that the $\mathrm{D_{4000}}$, with a sensitivity timescale of 1~Gyr, is not suitable for testing density wave theory since 1~Gyr is long enough to allow stars to travel between two spiral arms.
Overall, while numerous studies have explored age gradients in nearby galaxies, the conflicting results suggest that no single theory can fully explain the formation and evolution of all spiral features in the local Universe.

Spiral arms are not solely responsible for the azimuthal distributions observed in galaxies.
Gas and stellar migration also contribute significantly to shaping these variations.
For instance, \citet{Sanchez-Menguiano_2016} directly measure streaming motion in NGC~6754, revealing metal-rich gas moving outward along the trailing edge of the spiral arms, which induces azimuthal variations in metallicity.
Additionally, environmental effects -- accretion of gas and stars from fly-by satellites and/or mini-mergers -- can perturb the gravitational potential of the disc, redistributing gas and stars both radially and azimuthally without destroying the spiral structures \citep{Block_2002, Fraternali_2008, Eliche-Moral_2011}.

Measuring the radial migrations \citep{Roskar_2012, Sanchez-Menguiano_2016, Ruiz-Lara_2017}, classifying fly-by satellites \citep{van_Dokkum_2023, Zaritsky_2023}, and identifying past merging events in galaxies \citep{George_2018, Ristea_2022, Kelkar_2023} remain a significant observational challenge.
Simulations provide a complementary avenue to address these limitations, with many successfully reproduced spiral structures \citep{Sellwood_2011, Fujii_2011, Grand_2012, Donghia_2013,Grand_2014}.
Several studies on simulations \citep[e.g.,][]{Baba_2013, Grand_2016} find large-scale radial migrations in star particles driven by spiral arms in isolated disk galaxies.
Using the FIRE-2 hydrodynamic simulation model, \citet{Orr_2023} further demonstrate that spiral arms act as highways for radial migration of gas and metals.
\cite{Khoperskov_2018} on the other hand conclude from $N$-body simulations that vertical metallicity gradients may play a more important role in shaping the azimuthal variations in stellar metallicities compared with radial gradients.
Simulations offer comprehensive insights into a host galaxy’s environment by constructing merger trees using subhaloes \citep{Tweed_2009, Robles_2022}. These allow us to track the timing of merger events, the trajectories of fly-by satellites, and the properties of companions, such as their gas masses (\Mgas). 
This environmental context allows the comparison of spiral galaxies across diverse environments, as well as within a single galaxy before and after a merger event.
For one example, \citet{Pettitt_2017} simulate a multi-arm spiral galaxy that evolves into a two-arm spiral galaxy 400~Myr after introducing a perturbation by a companion satellite.
Their study highlights distinct behaviours in the bridge and tail, focusing on the spatial offset between gas and stellar arms, an outcome misaligned with predictions from both density wave theory and dynamic spiral theory.

The connection between spiral features and the azimuthal distribution of gas and stars remains a topic of debate, with one of the least explored questions being the long-term evolution of these distributions in spiral galaxies.
The motivation of this work is to investigate the long-term fluctuations of azimuthal variation in stellar age using the Auriga simulations, \red{without implementing rigid spiral density waves.}
Building on the work of \citet{Grand_2016}, we use a higher mass resolution dataset spanning a longer period.
We specifically focus on the role of environmental effects, such as mergers and fly-bys, in driving changes to age distributions.
We begin by introducing the Auriga simulations (Sec.~\ref{sec:auriga}) and our method to define the spiral arms in each snapshot (Sec.~\ref{sec:arm_def}).
We explore the environment of the host galaxy from merger tree data (Sec.~\ref{sec:merger}), as well as quantify the azimuthal variation at each snapshot (Sec.~\ref{sec:fno_dmu}).
In Sec.~\ref{sec:dis}, we discuss the connection between fluctuation in age gradient across spiral arms and the properties of ongoing mergers and fly-bys.
\red{We also compare to observational results from the literature to provide context.
We emphasise that this study focuses on how spiral features collectively respond to fly-by and merger events.
We do not introduce observational effects to generate mock observations and therefore only present qualitative instead of robust quantitative comparison.}
In Sec.~\ref{sec:conclu}, we summarise our conclusions of this work.
This work assumes a Hubble constant of $H_0$ = 100 $h$ km s$^{-1}$ Mpc$^{-1}$, where $h$ = 0.6777, and adopts cosmological parameters of $\Omega_\mathrm{m} = 0.307, \Omega_\mathrm{b} = 0.048, \Omega_\mathrm{\Lambda} = 0.693$ \citep{Planck_2014}.

\section{Methods}
\subsection{Auriga simulations}\label{sec:auriga}
\subsubsection{Auriga Level 3 dataset}\label{sec:aurigaL3}
To answer the scientific questions raised before, in this work, we use data products from the Auriga cosmological zoom-in simulations \citep{Grand_2017}, due to its high mass resolution ($\sim6\times10^3$M$_{\odot}$ in level 3) and the accessible individual halo merger and flyby histories.
Auriga is a follow-up to the ``Aquarius'' project of zoom-in simulations \citep{Marinacci_2014} and is run on the moving mesh magneto-hydrodynamics code {\sc arepo} \citep{Springel_2005,Pakmor_2016}.
The complete Auriga suite is comprised of \revi{30} haloes selected from the ``Evolution and Assembly of GaLaxies in their Environment'' (EAGLE; \citeauthor{Schaye_2015} \citeyear{Schaye_2015}) parent $(67.8~{\rm Mpc}/h)^3$ volume\ignorespaces
\footnote{\ignorespaces
We note that the original EAGLE simulation was run using a heavily modified version of {\sc gadget-3} called {\sc anarchy} (see \citeauthor{Schaye_2015} \citeyear{Schaye_2015}).
The resimulation of the Auriga haloes, however, was done in {\sc arepo} \citep[see][for complete details on this process]{Grand_2017}. 
}.
\revi{These haloes were randomly selected from 174 systems} that are approximately Milky Way mass -- $1\times 10^{12} \mathrm{M_\odot} < M_\mathrm{200} < 2\times 10^{12} \mathrm{M_\odot}$ -- at $z=0$ as well as are relatively isolated at $z=0$ (for computational reasons).
Once these targets were selected, particles within the radius of $4R_{200}$ (where \Rvir\ is the radius at which the mean enclosed mass volume density becomes 200 times the critical density) were resimulated at higher resolution using the method outlined in \cite{Jenkins_2010}.

The Auriga galaxy evolutionary model builds upon the Illustris model \citep{Vogelsberger_2014a,Vogelsberger_2014b,Genel_2014,Torrey_2014} and implements a wide range of (astro-)physical processes including, but not limited to: gravity, star formation, stellar feedback, magnetic fields, chemical enrichment, as well as black hole growth and feedback.
Briefly, the ISM's behaviour is described by the \cite{Springel_Hernquist_2003} effective equation of state.
This (and, indeed, all) equation(s) of state provide a physically motivated prescription for the dense, unresolved ISM.
In the \cite{Springel_Hernquist_2003} model, new star particles form stochastically beyond a threshold density of $n_{\rm H} \geq 0.13~{\rm cm}^{-3}$.
The star particles are treated as a single stellar population drawn from a \cite{Chabrier_2003} initial mass function and adopt the metallicity from the gas in which they form.
As they evolve, star particles return their mass and metals to the ISM via asymptotic giant branch winds as well as type Ia and II supernovae, with the metal yields from these events taken from \cite{Karakas_2010}, \cite{Thielemann_2003}, and \cite{Portinari_1998}, respectively.

Our analyses are carried out on the public data release of Auriga\footnote{\href{https://wwwmpa.mpa-garching.mpg.de/auriga/data.html}{https://wwwmpa.mpa-garching.mpg.de/auriga/data.html.}} \citep{Grand_2024}.
The data release of the Auriga simulations contains several haloes simulated at two different resolution levels: ``level 4'' (L4) with baryon mass resolution of $5\times10^4$ M$_\odot$ and temporal resolution of $\sim$160~Myr; and ``level 3'' (L3) with baryon mass resolution of $6\times10^3$ M$_\odot$ and temporal resolution of $\sim$300~Myr.
Table~\ref{tab:info} summarises all L3 haloes from the ``original'' sample in the public data release (we refer the reader to Table 2 of \citeauthor{Grand_2024} \citeyear{Grand_2024} for the complete details of 30 haloes in Auriga simulations).
This work is carried out on the L3 dataset since the higher mass resolution allows us to study the stellar populations in the interarm regions with higher fidelity and obtain a more accurate definition of the spiral arms.
Appendix~\ref{appendix:L3vsL4} provides a detailed comparison between spiral structures at L3 and L4 resolutions.
The public L3 dataset includes six spiral galaxies: Halo~6, 16, 21, 23, 24, and 27 (although we do not analyse Halo 6; see Section~\ref{sec:arm_def} and Appendix~\ref{appendix:halo6}).

To visualise the particle data, we rotate the galaxy face-on by calculating the unit vectors of angular momentum of star particles within 0.1\Rvir\ of the main halo.
We rotate the galaxy such that this angular momentum vector points in the $+\hat{z}$ direction to construct face-on images of galaxies.
Our analysis only involves star particles in the main halo disc within a box size of $20 \times 20 \times 5$ kpc$^3$, excluding halo stars \citep{Li_2024}. 
Furthermore, we limit the $z$-direction velocity to within $\pm$20 km/s, removing halo stars that may transiently appear on the disc plane during specific snapshots.
To create ``images'' of the system, we first map the discrete particle distributions of Auriga onto a 3D grid using a traditional smoothing method with a cubic spline kernel.
We convert this 3D grid into a 2D projection by integrating over the $\hat{z}$ direction (as defined by the angular momentum vector, see above) on a $300\times 300$ pixel grid in the $xy$ plane.
Following this method, we construct the young star mass map in Sec.\ref{sec:arm_def} and the mass-weighted age map in Sec.\ref{sec:fno_dmu}.

\begin{table*}
    \centering
    \begin{tabular}{c|rrrrrr}
    \hline
         (1) Halo &  (2) log(M$_\mathrm{200}$/ M$_\odot$) & (3) $R_\mathrm{200}$ (kpc) & (4) log(M$_*$/M$_\odot$) & (5) Morphology & (6) \# of mergers & (7) \# of close interactions\\
         \hline
         6 & 12.006 & 211.834 & 10.806 & Barred & 0 $^{\mathrm{b}}$ & 1 $^{\mathrm{b}}$ \\
         16 & 12.177 & 241.530 & 10.959 & Large disc, barred & 1 & 2\\
         21 & 12.151 & 246.688 & 10.944 & Unbarred & 0 & 3\\
         23 & 12.177 & 241.501 & 10.954 & Large disc, barred & 2 & 0\\
         24 & 12.167 & 239.568 & 10.939 & Large disc, unbarred & 1 & 3\\
         27 & 12.230 & 251.400 & 10.995 & Large disc, barred & 1 & 2\\
         \hline
    \end{tabular}
    \caption{Summary information of six haloes in the Original Milky Way-mass Auriga L3 simulations. The columns list: (1) the halo name; (2) the total mass inside $R_\mathrm{200}$; (3) $R_{200}$ is the radius at which the density becomes 200 times the critical density; (4) the stellar mass of the central galaxy; (5) a summary of the morphology of each halo; (6) number of mergers in the past 5~Gyr, including all mergers with stellar mass ratios greater than 1:100; (7) number of close interactions $^{\mathrm{a}}$ in the past 5~Gyr.\\
    Note.\\
    $^{\mathrm{a}}$ Close interactions are defined as encounters within $R_\mathrm{200}$ involving satellites with stellar mass exceeding 1\% of the host halo's stellar mass.\\
    $^{\mathrm{b}}$ Halo~6 is excluded in this study (see Sec~\ref{sec:arm_def} and Appendix~\ref{appendix:halo6}).
    }
    \label{tab:info}
\end{table*}

\subsubsection{Merger Trees}\label{intro:mergertree}

Gravitationally-bound structures in the Auriga simulations are identified using {\sc subfind} \citep{Springel_2001}.
Merger trees are built on top of the {\sc subfind} haloes using the procedure outlined in \cite*{Springel_DiMatteo_Hernquist_2005}.
These merger trees identify structures across different snapshots and enable direct tracking of galaxies over cosmic time.
The merger tree properties are also a part of the Auriga public data release. 
We refer the reader to \cite{Grand_2024} for complete details regarding the implementation and structure of merger trees.

\subsection{Classification of spiral arms, leading and trailing edge}\label{sec:arm_def}
\begin{figure}
    \centering
    \includegraphics[width=1\linewidth]{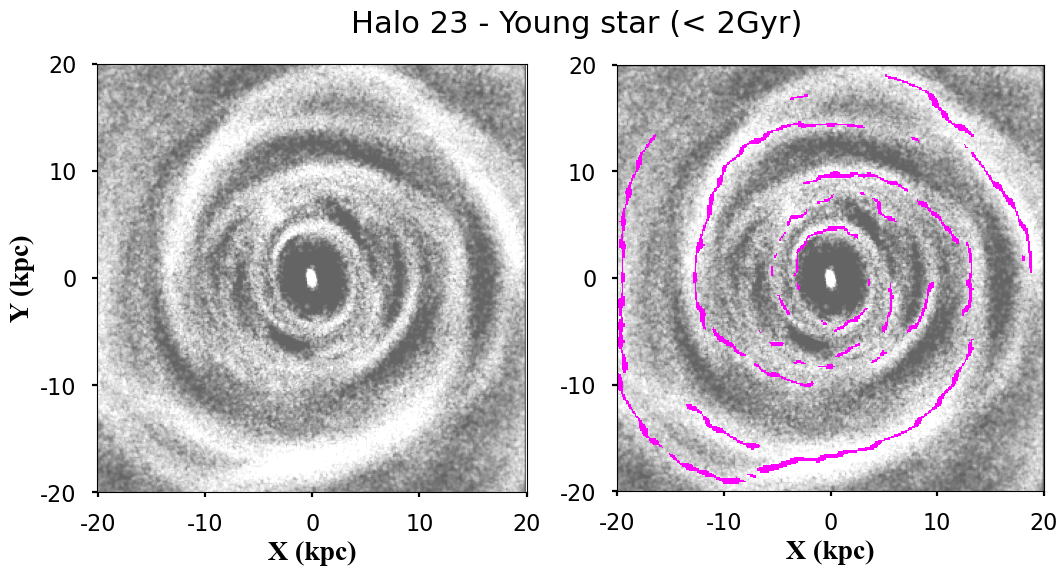}
    \caption{Young star (age $<2$~Gyr) mass map of Halo~23 at $z = 0$. The spiral arms identified by the algorithm (Sec.~\ref{sec:arm_def}) are presented as purple pixels in the right panel.}
    \label{fig:starmap}
\end{figure}

\begin{figure*}
    \centering
    \includegraphics[width=0.9\linewidth]{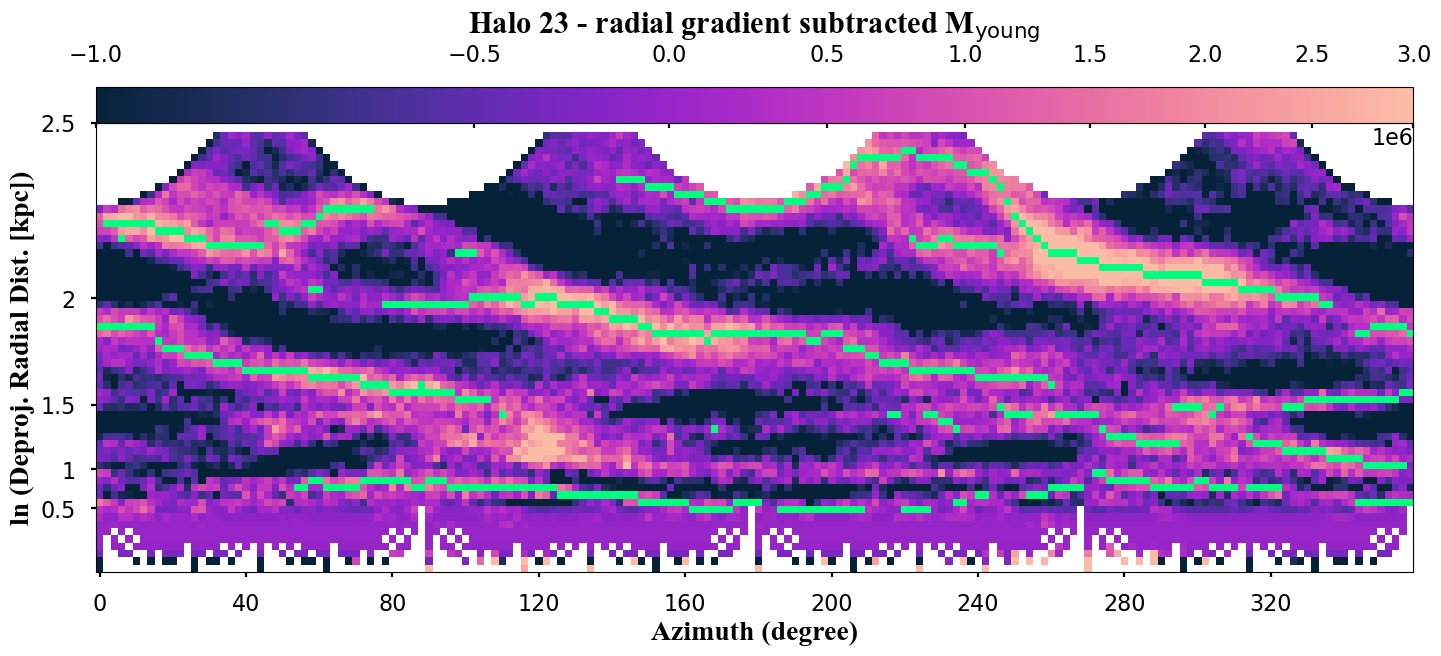}
    \caption{Phase diagram of Halo~23 at $z = 0$, colour-code by the young star ($<2$~Gyr) mass after subtracting the radial gradient. We use the moving average of each 3-pixel bin to represent the radial gradient. 
    The x-axis is the azimuth while the \revi{y-axis shows the logarithm of the radial distance (in units of kpc)}. An ideal spiral arm following Eq.~\ref{eq:arm1} is a straight line in the phase diagram. 
    We use an automatic algorithm, ridgeline walking, to identify pixels on the spiral arms, shown as green lines.}
    \label{fig:phase}
\end{figure*}

\begin{figure*}
    \centering
    \includegraphics[width=0.97\linewidth]{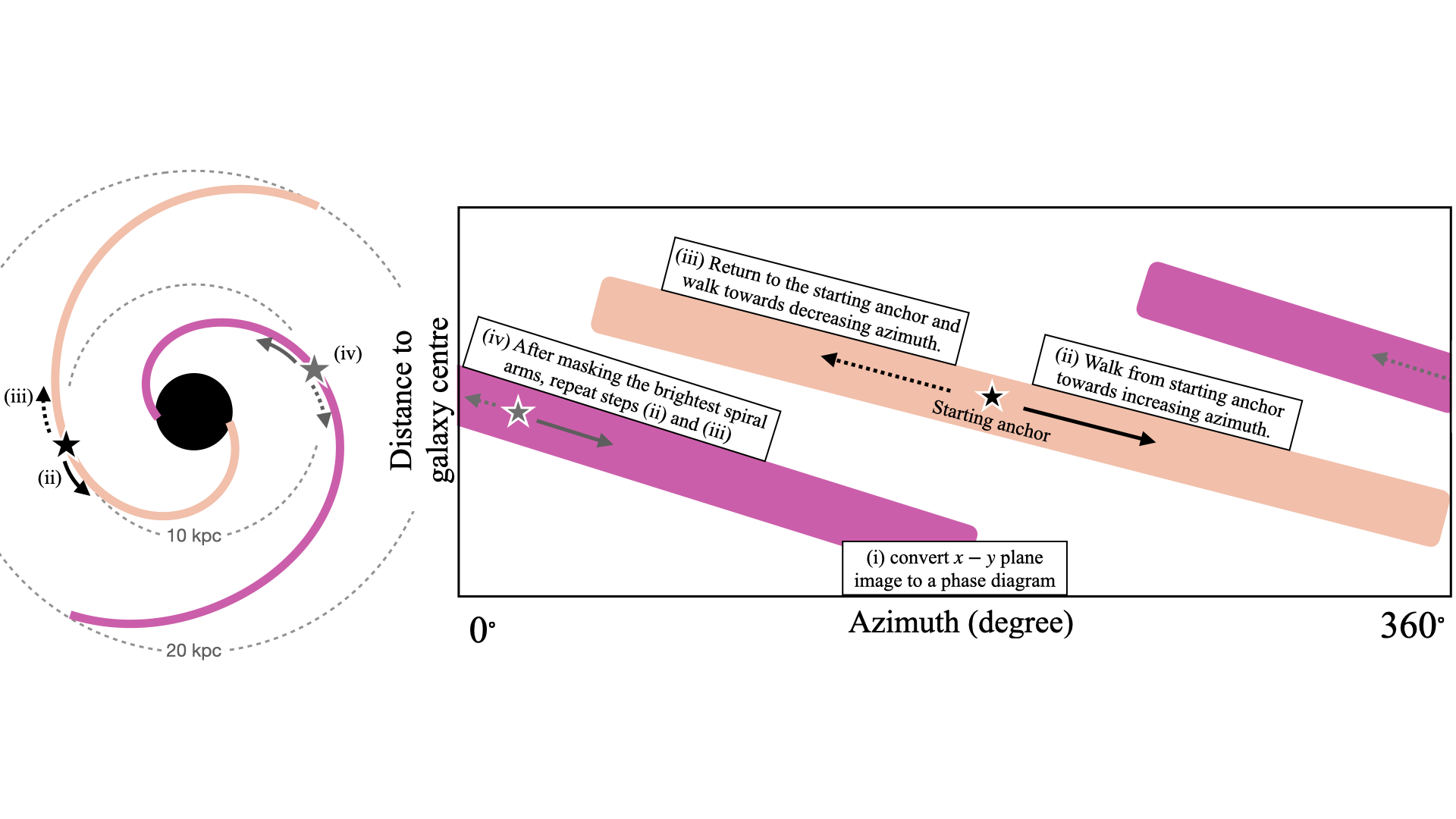}
    \caption{This schematic illustrates the four-step process of the ridgeline walking algorithm used to automatically identify spiral arms.
    The algorithm locates spiral arms on an azimuth $-$ radial distance diagram, where colour represents the young star mass after subtracting the radial gradient. 
    The starting anchor, marked by a black star symbol, is set at the brightest pixel in the middle radius (10kpc $\pm$ 1kpc). From this anchor, the algorithm searches for the next localized maximum and moves toward increasing azimuth, then returns to the anchor to walk towards decreasing azimuth, outlining the brightest spiral arm pixels. 
    After masking out the brightest spiral arm, the algorithm repeats this process to identify the second brightest spiral arm. 
    Appendix~\ref{sec:all_images} presents the young star maps for all five halos over the past 5Gyr, with spiral arm definitions overlaid (red lines).}
    \label{fig:algorithm}
\end{figure*}

\red{Various methods have been used to define spiral arm ridges, with many previous studies -- including simulations and observations -- tracing the young stars masses or fluxes~\citep[e.g.,][]{Dobbs_2014, Ho_2017, Shabani_2018, Pettitt_2020, Kreckel_2020, Silva-Villa_2022, Chen_2024b}.
Other studies have traced spiral arms using molecular gas clouds \citep[][]{Querejeta_2021}.
Slight spatial offsets can exist when spiral arms are traced at different wavelengths \citep{Yu_2018}.
The aim of this study is to examine how environmental effects influence the star formation in spiral galaxies. Since young stars closely trace recent star formation activities,}
we define spiral arms as regions of concentrated young stars by locating mass peaks in the young star mass (\Myoung; age $<2$~Gyr) maps, e.g., Halo~23 at $z$ = 0 in Fig.~\ref{fig:starmap}.
We design an automated algorithm, named \textit{``ridgeline walking''}\footnote{This algorithm is available on GitHub \href{https://github.com/Qian-HuiChen/ridgeline_walking}{https://github.com/Qian-HuiChen/ridgeline\_walking}.}, to identify the spiral arms in the Auriga L3 data.
Our algorithm aims to identify all pixels on the spiral arm ridge lines.
Although analytic spiral arms can be expressed as a formula \citep{Ho_2017,Chen_2024b}:
\begin{equation}
    r(\phi) = r_0 e^{\mathrm{tan\theta_p}(\phi-\phi_0)}, \label{eq:arm1}
\end{equation}
the ridgeline walking algorithm does not parameterise the spiral arms, as the spiral arms may not follow the analytic formula (Eq.~\ref{eq:arm1}) during interaction/merger events (e.g., 2.97~Gyr snapshot in Fig.~\ref{Afig:halo23}). 
This algorithm is carried out in four steps (Fig.~\ref{fig:phase} and Fig.~\ref{fig:algorithm}):
\begin{enumerate}[i]
    \item Firstly, we convert the face-on young star mass map to a phase diagram (azimuth versus distance; Fig.~\ref{fig:phase}). 
    To increase the contrast between spiral arms and inter-arms, we subtract the \Myoung\ by the radial gradient, represented by a moving mean value in every 3 pixels \revi{(0.2~kpc)}. 
    Fig.~\ref{fig:phase} is colour-coded by the \Myoung\ residuals, where the spiral arms are highlighted as positive values.

    \item Now, we locate the brightest pixel ($\phi_0$, $R_0$) within 9~kpc $< R <$ 11~kpc on the phase diagram, which works as the starting anchor for the ridgeline walking. 
    We apply the {\sc scipy.signal.find\textunderscore peaks} module \footnote{The detection threshold for peak detection is set to half of the median value of all points in the phase diagram. If the first signal detection fails, the algorithm will attempt a second threshold, set to half of the 25th percentile of all values in the phase diagram.} on a small region of the phase diagram, $\phi_0$ to $\phi_0+2^\circ$ and $R_0\pm2$~kpc, to detect a localised peak ($\phi_1$, $R_1$). 
    The algorithm extends the ridgeline to this new point ($\phi_1$, $R_1$) and detects the next localised peak within the region of $\phi_1$ to $\phi_1+2^\circ$ and $R_1\pm2$~kpc. 
    \revi{By continuously extending the ridgeline from $\phi_0$ to 360$^\circ$ and then wrapping to 0$^\circ$ \footnote{The pixel at ($\phi_i=360^\circ, R_i$) is identical as the one at ($\phi_i=0^\circ, R_i)$.}, and continuing to 360$^\circ$}, the algorithm classifies the inner ridgeline of the brightest spiral arm.

    \item To identify the outer ridgeline of the brightest spiral arm, the algorithm returns to the starting anchor ($\phi_0$, $R_0$). 
    Peak detection is operated in a new region of $\phi_0-2^\circ$ to $\phi_0$ and $R_0\pm$2~kpc, and continues in the direction of decreasing $\phi$.
    The algorithm completes the outer ridgeline of the brightness spiral arm by walking from $\phi_0$ to 0$^\circ$ and then from 360$^\circ$ to 0$^\circ$.

    \item The previous process identifies pixels on the brightest spiral arm and any connected spiral arms. 
    To locate spiral arms that are not connected to the brightest one, we mask out the spiral pixels identified in steps (ii) through (iii), spanning 2~kpc along the $y$-axis and 10$^\circ$ along the $x$-axis on the phase diagram. 
    The algorithm then \revi{repeats steps (ii) and (iii) until reaching a radius of 20~kpc}, and ultimately returns pixels on all spiral arm ridgelines.
\end{enumerate}

An example of the outcome of the ridgeline walking algorithm is shown as purple lines in the right panel of Fig.~\ref{fig:starmap} (Halo 23 at snapshot $z =$ 0) and Fig.~\ref{fig:phase}.

The main science goal of this work is to study the impact of environmental interactions on azimuthal variations in stellar age.
It is necessary to quantitatively separate the disc region into the leading and trailing edges, after identifying the spiral arm pixels.
Following \citet{Chen_2024}, we measure the parameter \dphi, i.e., azimuthal distance to the nearest spiral arm at the same galactocentric distance.
Given a pixel $T$, 
\begin{equation}
    |\Delta \phi_T| = \mathrm{min}(|\phi_T - \phi_{arm1}|, |\phi_T - \phi_{arm2}|, ..., |\phi_T - \phi_{armN}|),
\end{equation}
where pixel $arm1$, $arm2$, ..., and $armN$ are on the spiral arm ridgelines and at the same galactocentric distance as pixel $T$.
We assign a positive (negative) \dphi\ when the targeted pixel is on the trailing (leading) edge of the nearest spiral arm.
We perform the spiral arm identification algorithm, followed by the \dphi\ measurement, to snapshots from $z = 0$ back to $z = 0.46$ (lookback time of 5~Gyr) when the disc became stable and spiral arms became prominent.
We find that in Halo~6, the spiral features are tightly wound, nearly ring-like and not prominent in the disc (Appendix~\ref{appendix:halo6}).  These characteristics make it more difficult to identify spiral arms and result in fewer pixels available for us to statistically compare the leading and trailing edges.
Consequently, we focus our analysis on the other five L3 halos and exclude Halo~6.
Appendix~\ref{sec:all_images} presents the young star maps of all our samples over the past 5~Gyr, overlaid with the defined spiral arms.

We note that there are various other methods by which spiral arms are identified in simulations and observations. 
Some observational work \citep{Ho_2018, Querejeta_2021} identify spiral arms by visual inspection on images or phase diagrams.
Other studies first select the main spiral arm regions by eye and then fit the areas with the analytic formula (Eq.~\ref{eq:arm1}).
The large number of snapshots in our sample, however, makes it impractical to classify the spiral arms manually, necessitating an automatic identification procedure.
\cite{Davis_2014} developed an automated detection of spiral arm segments based on over 600,000 objects in the SDSS -- SpArcFiRe.
This algorithm is successfully applied to two spiral galaxies at $z\sim$ 0.3 \citep{Chen_2024}.
However, designed for observation images, SpArcFiRe does not perform well on the Auriga-simulated galaxy images, potentially due to the absence of observational effects such as the point spread function.
Another software by \citet{Forgan_2018} classifies the spiral structures in a protostellar disc in hydrodynamic simulations.
The tensor classification in \citet{Forgan_2018} is designed to operate on smooth continuous fields.
However, the spiral arms are disrupted and do not follow the analytic formula during environmental interactions.
Since one of our science goals is to compare the spiral galaxies before and after interactions with satellites, we do not adopt the algorithm from \citet{Forgan_2018} in this work.

\section{Analysis and results}
\subsection{Quantifying the azimuthal variation in stellar age}\label{sec:fno_dmu}
\begin{figure}
    \centering
    \includegraphics[width=0.98\linewidth]{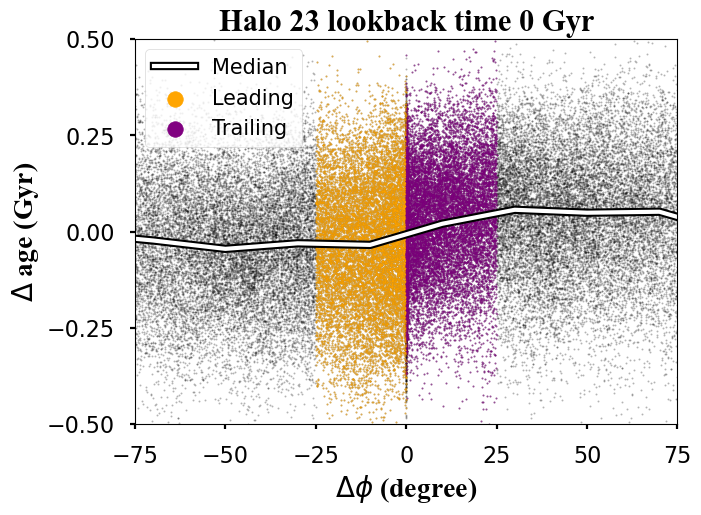}
    \caption{The parameter \dage\ (stellar age with the radial gradient subtracted) varies with the azimuthal distance \dphi\ to the spiral arms. 
    Spaxels on the leading (trailing) edge are assigned with negative (positive) \dphi.
    The region $-25^\circ < $ \dphi $ < 25^\circ$ is highlighted, where the influence of the spiral arms on the age pattern is most significant. 
    Only the coloured pixels within this range are used to quantify azimuthal variations at each snapshot (Fig.~\ref{fig:age_dist}).
    The solid white line represents the median value within each moving 20$^\circ$ bin.}
    \label{fig:dphi-dage}
\end{figure}

Generally, the distribution of young stars is expected to exhibit a closer association with spiral arms, whereas older stars tend to be more diffusely spread across the disc \citep{Grasha_2017}. 
To highlight the impact of spiral arms on stellar age, the following analysis focuses exclusively on young stars ($<2$~Gyr) in each snapshot. 
In observed galaxies, stellar age varies in both radial and azimuthal directions simultaneously, with the radial gradient \citep{Parikh_2021} generally being more prominent than the azimuthal variation \citep{Shabani_2018}.
To remove the radial trend of the stellar age, we subtract the average age of a moving 3-pixel (0.4~kpc) wide ring from the stellar age.
We denote the age residual as \dage.
Fig.~\ref{fig:dphi-dage} compares the \dage\ in the leading edge (\dphi $<0^\circ$) and the trailing edge (\dphi $>0^\circ$) of the spiral arms in Halo 23 at $z = 0$.
We present the moving medians of \dage\ of each 20$^\circ$ as a silver solid line in Fig.~\ref{fig:dphi-dage}.
We find lower \dage\ in the leading edge (\dphi $<0^\circ$), indicating more young stars in the leading edge than the trailing edge.
The younger leading edge is consistent with observations in the local Universe \citep[e.g., NGC~1566 in][]{Shabani_2018} and also aligns with the prediction of density wave theory, within the co-rotation radius.

\begin{figure*}
    \centering
    \includegraphics[width=0.43\linewidth, height=3.3in]{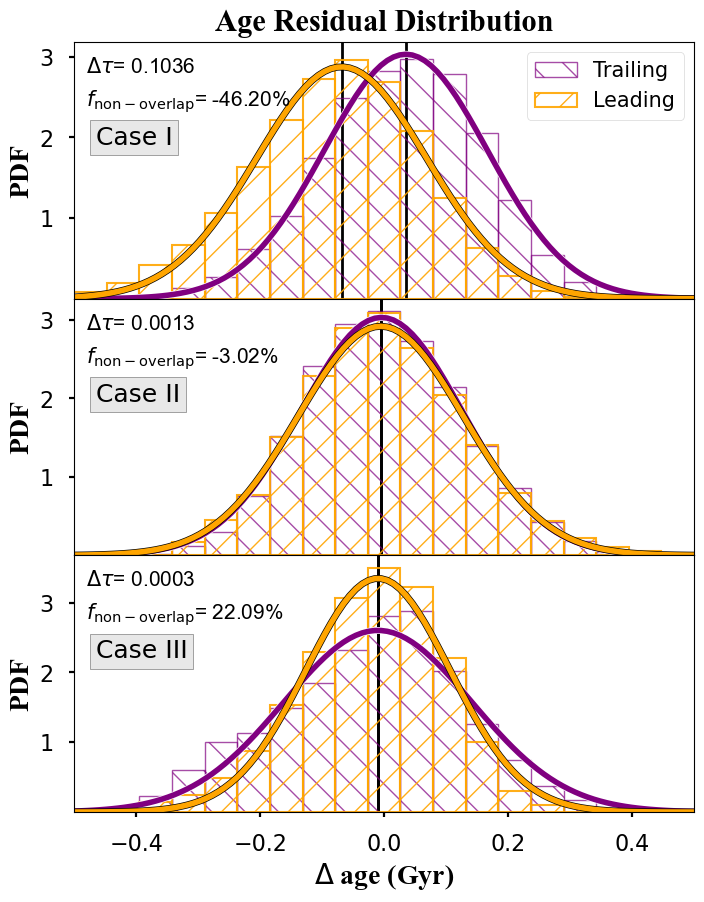}
    \includegraphics[width=0.43\linewidth, height=3.3in]{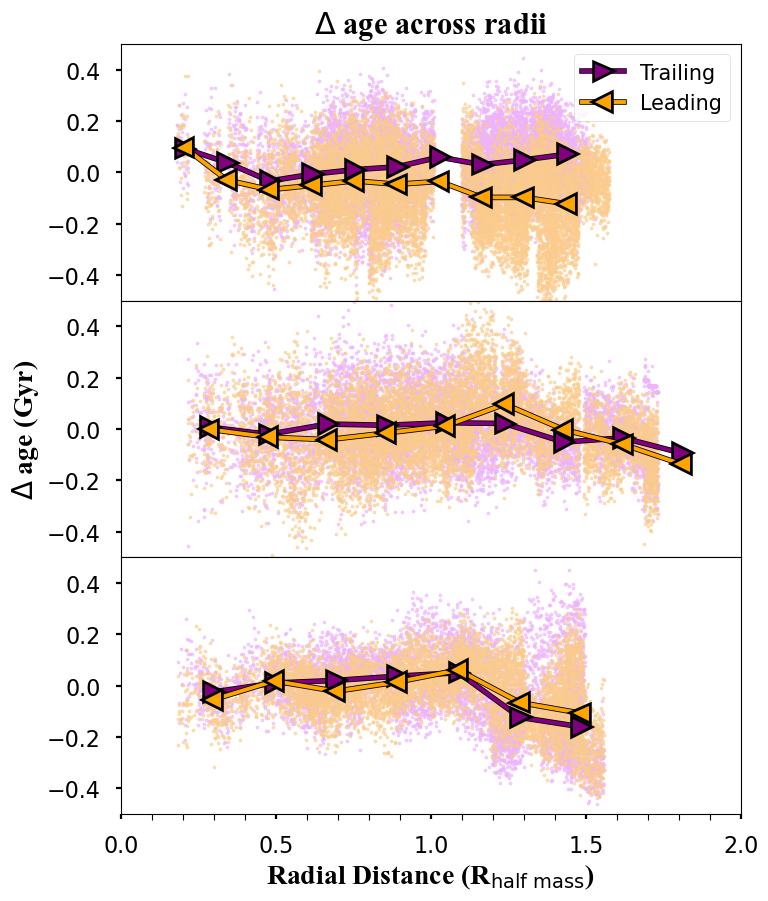}
    \caption{\textbf{Left:} Three cases of \dage\ distributions at the leading and trailing edges.
    \textbf{Case~\textsc{i}}: The \fno\ is large with a significant \dmu, indicating an evident age gradient across the spiral arms.
    \textbf{Case~\textsc{ii}}: Both \fno\ and \dmu\ are small, suggesting little to no azimuthal age variation.
    \textbf{Case~\textsc{iii}}: The \fno\ is large, but the \dmu\ is near zero, indicating distinct age distributions on each side of the spiral arms, with the difference mainly laying in the tail of the distribution.
    \textbf{Right:} Radial profiles of $\Delta$age (age residual after removing radial medians) for three cases, colour-coded by the location relative to the spiral arms. This panel illustrates how \fno\ and \dmu\ statistically represent the azimuthal variation in stellar ages. In Case~\textsc{i}, the leading edge (purple) is consistently older at most radii; in Case~\textsc{ii}, both sides exhibit similar ages; and in Case~\textsc{iii}, while the median ages are comparable, the trailing-edge distribution (purple scatters) is more dispersed.} 
    \label{fig:age_dist}
\end{figure*}

We quantify the age gradient across spiral arms at each snapshot.
This analysis allows us to trace azimuthal variations over cosmic time, especially before and after mergers.
We focus on star particles near the spiral arms (|\dphi|\ $<25^\circ$), where the spiral arms' influence is expected to be the primary factor (shown as coloured data in Fig.~\ref{fig:dphi-dage}).
Fig.~\ref{fig:age_dist} shows the distribution of \dage\ in the trailing (purple; $0^\circ < $\dphi $< 25^\circ$) and leading edge (orange; $-25^\circ < $ \dphi $ < 0^\circ$).
We fit these histograms with Gaussian distributions separately.
The azimuthal variations are quantified at each snapshot using two parameters: \dmu\ and \fno.
The first parameter, \dmu, is defined as the difference between the average ages of the two Gaussian distributions, 
\begin{equation}
    \Delta\tau = |\overline{\Delta{\rm age}}_{\rm trailing} -  \overline{\Delta{\rm age}}_{\rm leading}|.
\end{equation}
The second parameter, \fno, is the non-overlap fraction between the two Gaussian distributions, calculated as
\begin{equation}
    f_{\mathrm{non-overlap}} = 
    \begin{cases}
    (1 - \frac{\mathrm{overlap\ area}}{\mathrm{total\ area}}) \times 100\%, 
    &\overline{\Delta{\rm age}}_{\rm trailing} <  \overline{\Delta{\rm age}}_{\rm leading} \\
    - (1 - \frac{\mathrm{overlap\ area}}{\mathrm{total\ area}}) \times 100\%, 
    &\overline{\Delta{\rm age}}_{\rm trailing} > \overline{\Delta{\rm age}}_{\rm leading}
    \end{cases}
\end{equation}
We assign a negative \fno\ when the leading edge has a younger average \dage\ (\dmu\ $>0$), otherwise, a positive \fno\ is assigned.

In terms of the relative relations of \dmu\ and \fno, we identify the following three cases in Auriga L3 simulations.
\begin{itemize}
    \item Case~\textsc{i}: A significant age gradient across the spiral arms is expected to result in a large magnitude of |\fno| and |\dmu| in a given snapshot.
    \item Case~\textsc{ii}: A low magnitude of |\fno|\ with |\dmu|\ close to zero indicates no or little age azimuthal variation in the galaxy.
    \item Case~\textsc{iii}: A notable case is that a large |\fno| is observed but along with |\dmu|\ close to zero. 
    In this case, the average \dage\ remains similar for both populations, but the different broadness is responsible for the large |\fno| (no matter what sign it has). 
\end{itemize}
Together, \dmu\ and \fno\ robustly quantify the azimuthal age variations at each redshift.
We will discuss our findings of azimuthal age azimuthal in detail in Sec.~\ref{sec:age_merger_connect}.

\subsection{Merging and interaction time}\label{sec:merger}
Mergers and interactions can impact the azimuthal distribution of stars and gas in spiral galaxies, which may even induce or reshape the spiral arms \citep{Villalobos_2008}. 
In this section, we introduce our determination of merging events and fly-by companions by analysing the merger trees \citep[see Sec.~\ref{intro:mergertree} and][]{Grand_2024}.

\begin{figure*}
    \centering
    \includegraphics[width=0.42\linewidth]{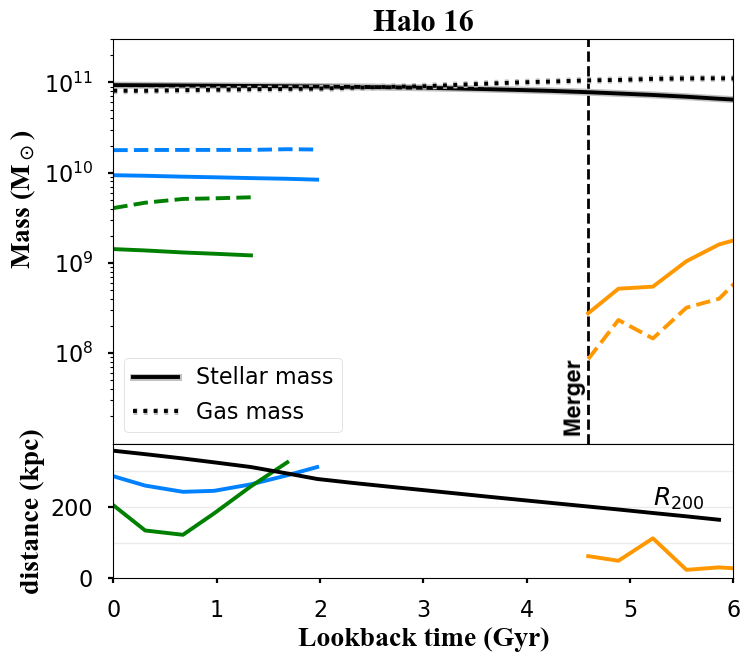}
    \includegraphics[width=0.55\linewidth]{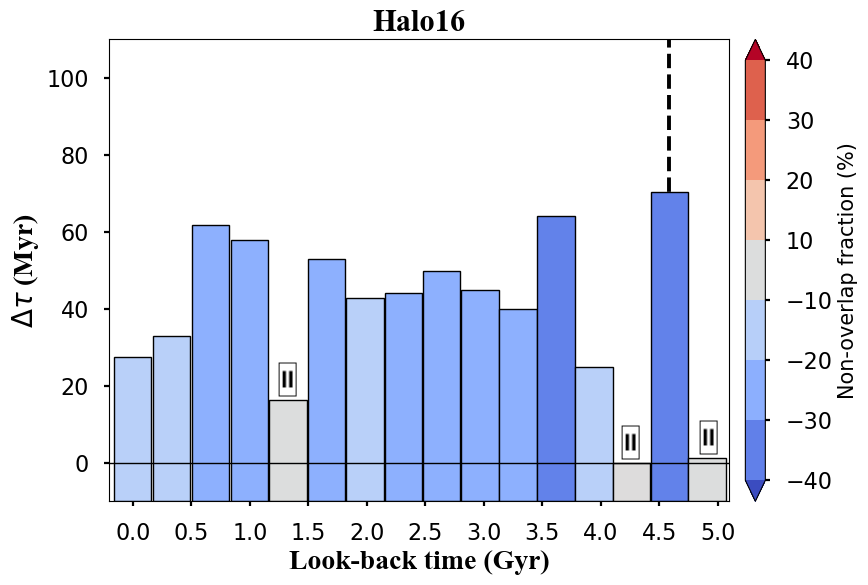}
    \caption{\textbf{Left:} Stellar mass (solid lines) and gas mass (dashed lines) of the main halo (black lines) and the nearby satellites (blue/green/orange/brown lines, if they exist) from the lookback time of 6~Gyr. We only present satellites within \Rvir\ of the main haloes and with a stellar mass larger than 1\% of the main halo. 
    Dashed vertical lines indicate merging events.
    The bottom panel presents the distance between the main halo and the satellites, with their corresponding colours. 
    \textbf{Right:} The evolution of \fno\ and \dmu\ over the past 5~Gyr, with merger events marked by dashed vertical lines. 
    The histogram is colour-coded by \fno\ for better comparison with Fig\ref{fig:merge_prop}, and the bar height represents \dmu. 
    Values of \dmu\ between $-10$ and $0$~Myr are shown at all times solely for visualisation, allowing the colour of small (near zero) \dmu\ to be seen.
    The snapshots aligning with \ctwo\ and \cthree\ are highlighted with a \textsc{ii} symbol or \textsc{iii} symbol on the top.
    Most snapshots exhibit a negative \fno\ and a large \dmu\ (\cone), indicating a younger leading edge compared to the trailing edge, a trend consistent across our entire sample.
    We identify a fly-by satellite (green line on the left) approaching Halo~16 around 1~Gyr ago, coinciding with a \ctwo\ snapshot 1.33~Gyr ago. 
    A satellite (orange line in the left) merged into Halo~16 at a lookback time of $\sim$4.5~Gyr, along with two \ctwo\ snapshots observed around the merging time.
    }
    \label{fig:result_16}
\end{figure*}

\begin{figure*}
    \centering
    \includegraphics[width=0.42\linewidth]{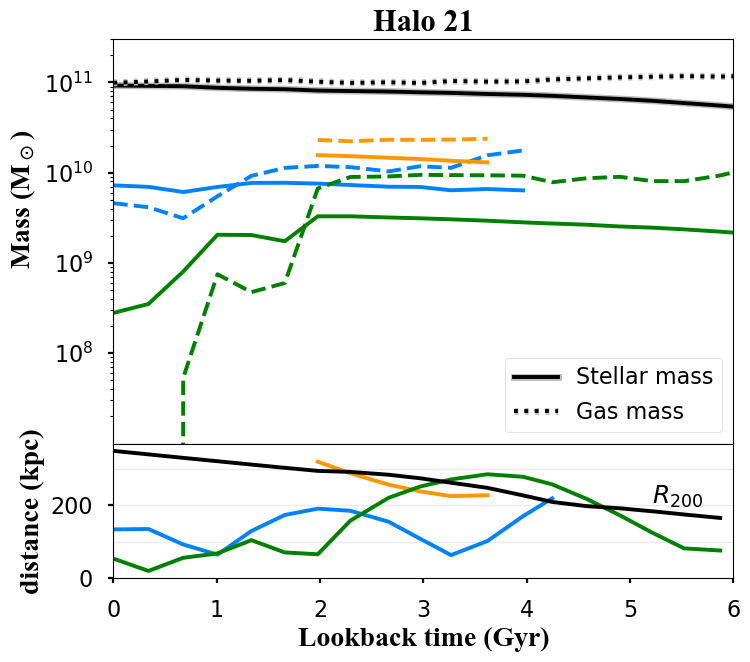}
    \includegraphics[width=0.55\linewidth]{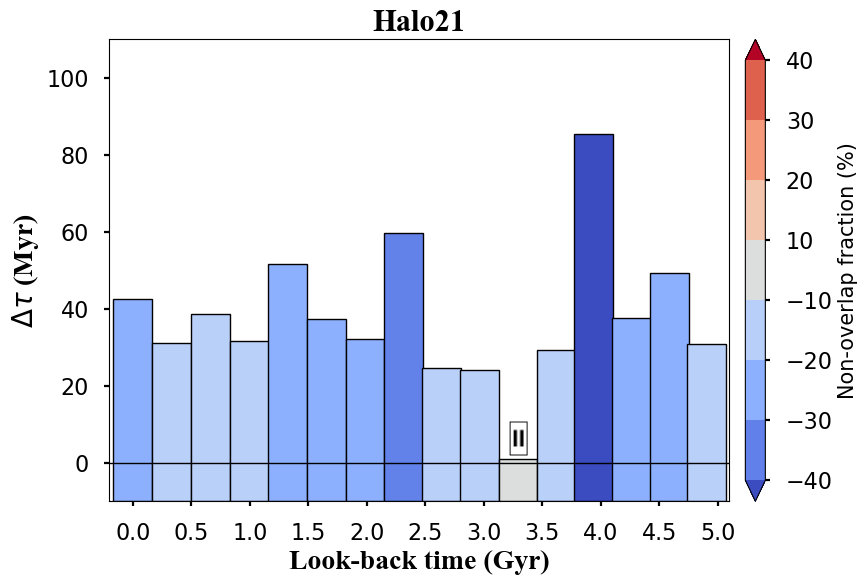}
    \caption{Same as Fig.~\ref{fig:result_16} but for Halo~21.
    We observe one \ctwo\ snapshot at a lookback time of 3.3~Gyr, coinciding with a fly-by satellite (blue line in the left panel).}
    \label{fig:result_21}
\end{figure*}

\begin{figure*}
    \centering
    \includegraphics[width=0.42\linewidth]{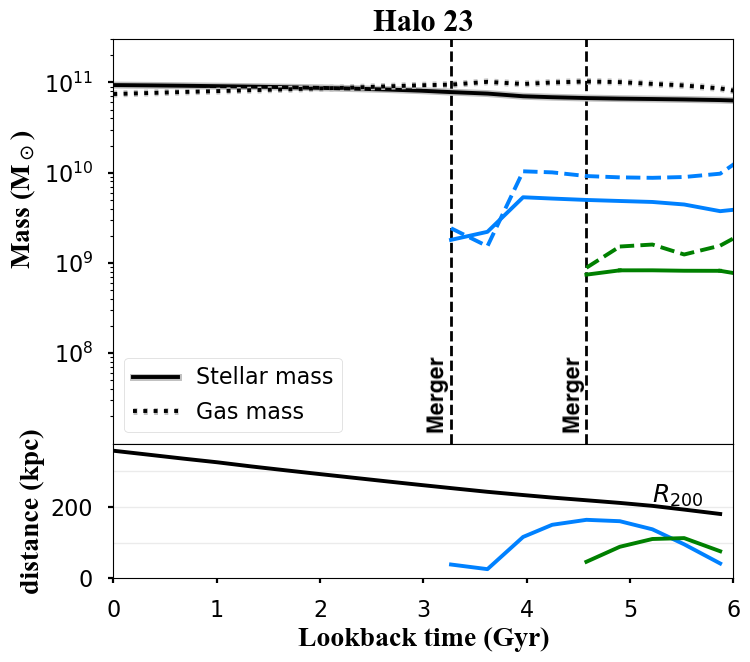}
    \includegraphics[width=0.55\linewidth]{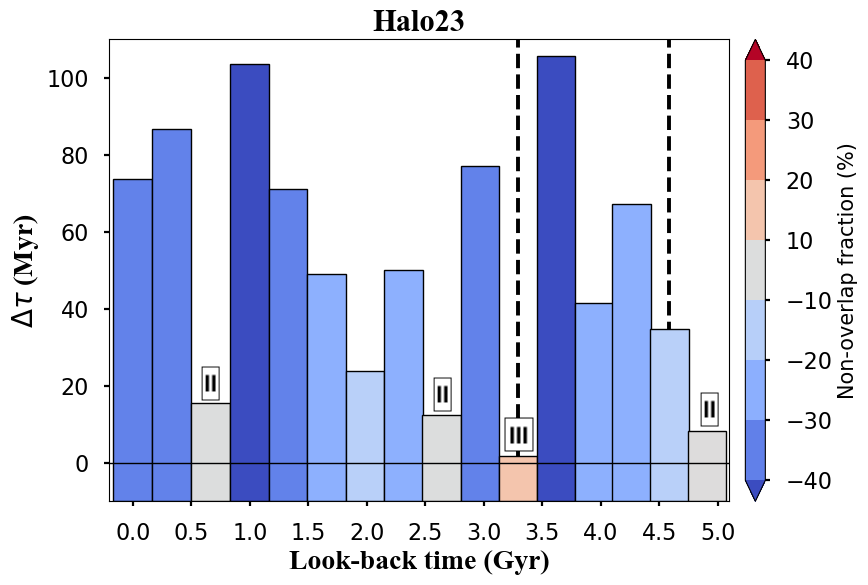}
    \caption{Same as Fig.~\ref{fig:result_16} but for Halo~23.
    We observe a \ctwo\ snapshot $\sim5$~Gyr ago, with a satellite (green line in the left panel) merging in $\sim300$~Myr.
    At around 3~Gyr ago, a \ctwo\ snapshot and a \cthree\ snapshot are observed, coinciding with a merging event (blue line in the left panel).
    We find a \ctwo\ snapshot at $\sim$0.6Gyr ago, with no sign of any satellite over 1\% stellar mass of Halo~23 within \Rvir.
    }
    \label{fig:result_23}
\end{figure*}

\begin{figure*}
    \centering
    \includegraphics[width=0.42\linewidth]{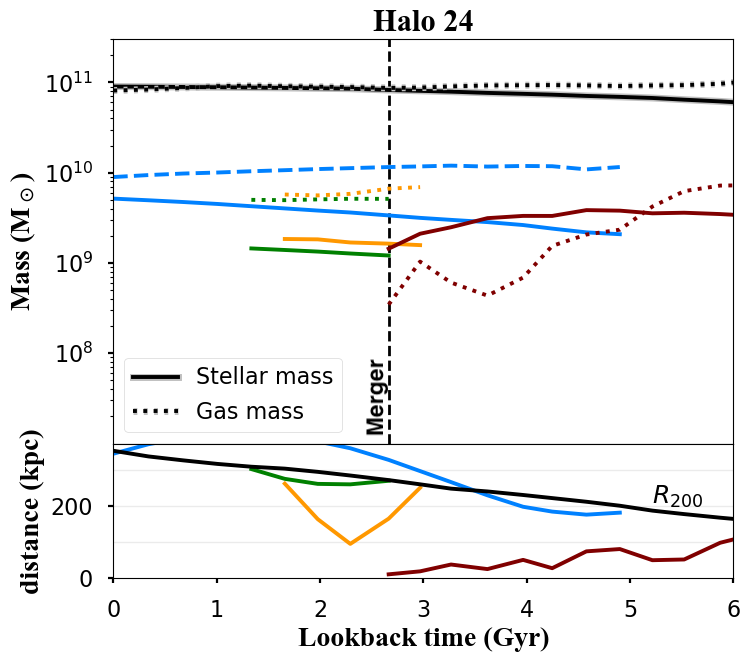}
    \includegraphics[width=0.55\linewidth]{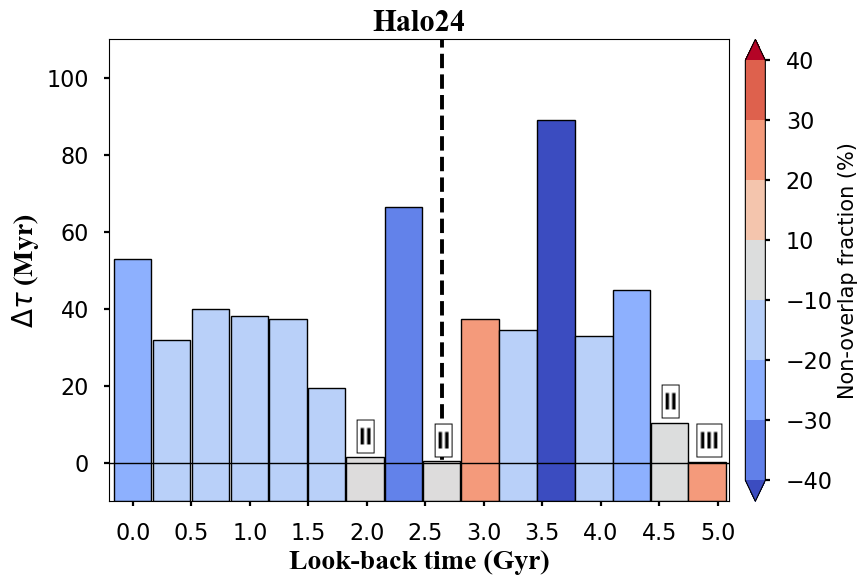}
    \caption{Same as Fig.~\ref{fig:result_16} but for Halo~24.
    Around 5~Gyr, we find one \ctwo\ snapshot and one \cthree\ snapshot, when the gas of a satellite (dark red line in the left panel) got stripped into Halo~24.
    When this satellite merged in, a \ctwo\ snapshot is found at a lookback time of $\sim2.6$~Gyr.
    Another \ctwo\ snapshot is identified at a lookback time of 2~Gyr, with a fly-by event happening (orange line in the left panel).
    }
    \label{fig:result_24}
\end{figure*}

\begin{figure*}
    \centering
    \includegraphics[width=0.42\linewidth]{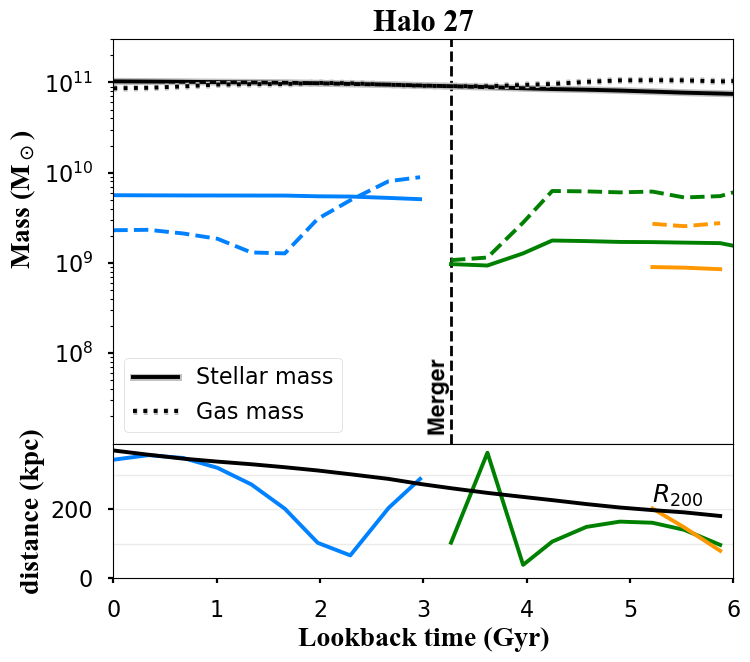}
    \includegraphics[width=0.55\linewidth]{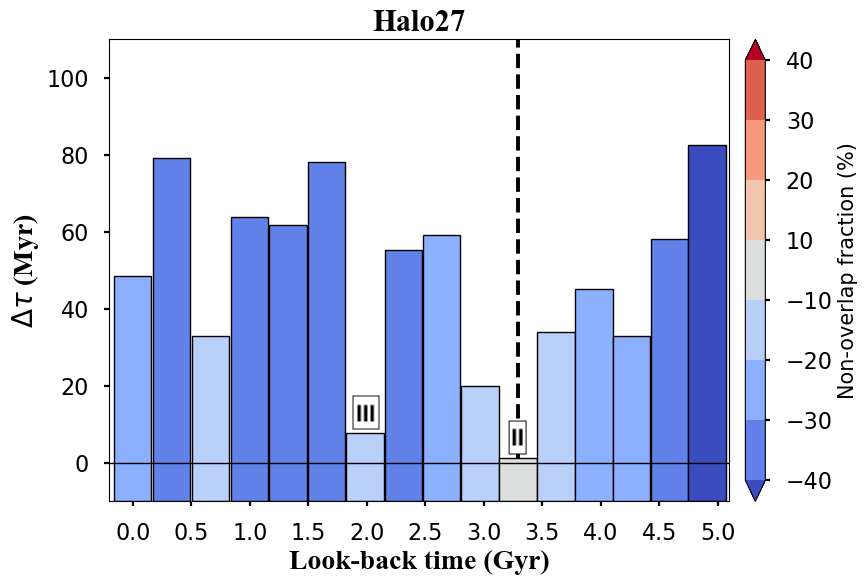}
    \caption{Same as Fig.~\ref{fig:result_16} but for Halo~27.
    Around 3.3~Gyr, we find a \ctwo\ snapshot, when a satellite (green line in the left panel) merged into Halo~27.
    A \cthree\ snapshot is found at a lookback time of $\sim2$~Gyr with a fly-by event happening (blue line in the left panel).
    }
    \label{fig:result_27}
\end{figure*}

We track the stellar mass of the main halo over the past 6~Gyr with solid lines and its gas mass with dotted lines in the left upper panel of Fig.~\ref{fig:result_16}, Fig.~\ref{fig:result_21}, Fig.~\ref{fig:result_23}, Fig.~\ref{fig:result_24} and Fig.~\ref{fig:result_27}.
We also show the mass of all satellites within \Rvir\ and with a stellar mass over 1\% of the main halo.
The bottom sub-panels demonstrate the distance from the satellites to the main halo (\Dsat).
When a subhalo is approaching the main halo (i.e., \Dsat\ is decreasing) and finally disappears at a snapshot, we determine that the merger happens at that snapshot, as highlighted by the dashed vertical lines.
The \Dsat\ is essential to distinguish merged satellites from those that move outside \Rvir, which shows an increasing \Dsat.
We can also identify fly-by events, when a satellite approaches and moves away from the main halo, based on the U-shape in \Dsat.

Halo~23 and Halo~27 live in relatively quiet environments, with only two companions within \Rvir\ over the past 5~Gyr.
Halo~16, 21, and 24, however, have encountered three or more sub-haloes within their \Rvir.
We find that most galaxies in our sample (except for Halo~21) experienced mergers over the past 5~Gyr, with stellar mass ratios between the merging galaxy and the main halo of $0.1-0.25$ ($0.01-0.1$), classified as minor (mini) mergers, respectively \citep{Remus_2022, Bottrell_2024}. 
We notice that \Mgas\ decreases along with \Mstar\ before merging, reflecting that the satellite's gas is being stripped by the main halo.
Fly-by events are also detected in all galaxies (except Halo~23), with stellar mass ratios in a similar range of $\sim1:10$ to $1:100$.
We will discuss the influences from both merger and fly-by events on the azimuthal variations in Sec.~\ref{sec:dis}.

\subsection{Age patterns before and after merging}\label{sec:age_merger_connect}
We apply the analyses of measuring age offset (Sec.~\ref{sec:fno_dmu}) and merging time (Sec.~\ref{sec:merger}) to snapshots from $z = 0$ back to $z = 0.46$ (lookback time of 5~Gyr), covering snapshots 63 to 48, respectively. 
The right panel of Fig.~\ref{fig:result_16}, Fig.~\ref{fig:result_21}, Fig.~\ref{fig:result_23}, Fig.~\ref{fig:result_24} and Fig.~\ref{fig:result_27} presents the age azimuthal variation over the past 5 Gyr of all haloes, with the dashed vertical lines indicating the timing of merger events.
To quantify the significance of the non-overlap fractions, we draw random samples of two identical Gaussian distributions, with a 10$^4$ sample size, 1000 times. 
The average \fno\ from the random sampling is 9.98\%, represented by the grey regime in the colour bar. 
When a galaxy's \fno\ falls within the grey regime, the age distributions on either side of the spiral arms can be considered statistically indistinguishable, therefore, no age azimuthal variation is present in this galaxy.

In the past 5~Gyr, our haloes, at most redshifts, follow \cone\ and show a negative \fno\ (tall blue histograms), referring to a younger leading edge than the trailing edge.
This situation aligns with the prediction of density wave theory when the new-borne stars overtake the density waves inside the co-rotation radius (Sec.~\ref{sec:intro}).
One snapshot (Halo~24 at 3~Gyr ago) is detected to follow \cone\ but a positive \fno\ (tall orange histogram in the bottom middle panel), indicating age azimuthal variation but a younger trailing edge.
During this time, there was a satellite (dark red line in Fig.~\ref{fig:result_24}) approaching Halo~24 and continuously losing a substantial fraction of its gas mass.
At a lookback time of 3 Gyr, 300~Myr (one snapshot) before this satellite fully merges with Halo~24, the spiral structure is highly disturbed, showing pronounced tidal features, including a strong stream surrounding the right side of the galaxy (Fig.~\ref{Afig:halo24}).  
This snapshot stands out from other redshifts of Halo~24 and is most likely driven by the prominent bridge structure and tidal disruptions caused by the merger.

We also find that a few snapshots of our galaxies follow \ctwo\ (short grey histograms), indicating little or no age azimuthal variation.
Around 50\% \ctwo\ (including lookback times of $\sim4.3$~Gyr and $\sim5$~Gyr in Halo~16, $\sim5$~Gyr in Halo~23, $\sim2.6$~Gyr in Halo~24, and $\sim3.3$~Gyr in Halo~27) takes place near or at the snapshot that a merger happens (dashed vertical lines in Fig.~\ref{fig:result_16}, Fig.~\ref{fig:result_21}, Fig.~\ref{fig:result_23}, Fig.~\ref{fig:result_24} and Fig.~\ref{fig:result_27}).
This finding suggests that gravitational perturbation and merged-in stars from the satellites destroy the age azimuthal variation in the main halo.
We notice that some merging events, i.e., $\sim4.6$~Gyr in Halo~16, $\sim2.6$~Gyr in Halo~24, and $\sim3.3$~Gyr in Halo~27, lead to more open (larger pitch angle) spiral structures and reduce the number of spiral arms, see Fig.~\ref{Afig:halo16}, Fig.~\ref{Afig:halo24} and Fig.~\ref{Afig:halo27}, respectively.
This supports the hypothesis that mergers reset the density wave potential in the disc, eliminating the age gradient across the spiral arms during the rebuilding process.

Interestingly, we observe a small difference in average age \dmu\ but a large non-overlap fraction \fno\ (short coloured histogram) at a lookback time of $\sim$ 3.3~Gyr in Halo~23, 5~Gyr in Halo~24, and 2~Gyr in Halo~27, consistent with \cthree. 
It is important to note that the sign of \fno\ is determined by whether the leading edge has a higher $\mu$ than the trailing edge. 
However, when \dmu\ is close to zero, the sign of \fno\ can be highly uncertain.
The combination of small \dmu\ and large \fno\ (\cthree) suggests that the age distribution differs between the leading and trailing edges, primarily driven by the distribution's broadness, while the average age remains similar on both sides of the spiral arms.
We find all three snapshots coincide with significant gas loss in the companion galaxy within 300~Myr (one snapshot).

After a merger, the \fno\ spends $\sim 600$ Myr ($\sim$1-2 snapshots) decreasing to around $-20$\% to $-40$\%, along with a larger \dmu\ compared to the merger snapshots, suggesting a recovery of azimuthal variations and the reformation of a younger leading edge. 
A higher temporal resolution simulation can impose more constraints on the timescale required for rebuilding the age gradient across spiral arms.
We will further discuss the three cases and possible impacts from the environment (mergers and fly-bys) in Sec.~\ref{sec:dis}.

\section{Discussion}\label{sec:dis}
In this section, we discuss the snapshots corresponding to three cases: \cone\ (high \fno\ and \dmu; Sec.~\ref{sec:case1}), \ctwo\ (low \fno\ and \dmu; Sec.~\ref{sec:case2}), and \cthree\ (high \fno\ and low \dmu; Sec.~\ref{sec:case3}), with a focus on the environmental impacts of mergers and fly-by events. 
We also discuss our findings from the Auriga simulations with observed trends reported in the literature (Sec.~\ref{sec:obs_sim}).

\subsection{Case \textsc{i}: Consistency with Density Wave Theory}\label{sec:case1}
Most of the snapshots in Auriga simulations align with \cone, showing a younger leading edge.
We identify our spiral arms based on the localised peak on the young star maps (Section~\ref{sec:arm_def}).
According to density wave theory, star formation is stimulated when gas clouds enter density waves, and subsequently stars overtake density waves within co-rotation radii, leading to a secondary concentration of young stars in the leading edge.
We therefore expect to find more young stars on the leading edge, and fewer young stars on the trailing edge in a galaxy undergoing density wave theory.
In most snapshots, we find a large negative \fno\ with \dmu\ $>10$~Myr, indicating a younger stellar population within the leading edge than the trailing edge.
This scenario aligns with the predictions of density wave theory inside the co-rotation radius.
\red{We further discuss how age patterns vary across radii in Appendix~\ref{appendix:radial_bin}, given that density wave theory predicts distinct azimuthal variations cross the co-rotation radius.}

Other physical mechanisms may also contribute and give rise to azimuthal variations on each side of the spiral arms, with radial migration being one of the most discussed in the literature.
In Auriga L4, \citet{Grand_2016} find that Halo~25 exhibits a faster peculiar azimuthal velocity and a peculiar inward radial velocity at the leading edge.
Since a negative metallicity radial gradient is observed, metal-poor particles in the outer disc migrate inward along the leading edge of the spiral arms.
Therefore, this radial migration leads to lower [Fe/H] in the leading edge.
The stellar velocity data in Auriga L3 can provide further constraints on stellar radial migrations and the influence on azimuthal age variation. 
This is especially true as the velocity field is altered during galaxy interactions, potentially contributing to changes in stellar age distribution. 
This topic will be expanded into a separate publication; our discussion in this work will focus on the predictions of density wave theory and the environmental effects on age asymmetry.

\subsection{Case \textsc{ii}: Implications on gas-rich interactions}\label{sec:case2}
We observe \ctwo\ with small \fno\ and small \dmu\ (Fig.~\ref{fig:result_16}, \ref{fig:result_21}, \ref{fig:result_23}, \ref{fig:result_24} and \ref{fig:result_27}) in all five haloes.
Most \ctwo\ snapshots occur near a fly-by or merger event, except for Halo~23 at 0.7~Gyr ago.

\begin{figure*}
    \centering
    \includegraphics[width=0.4\linewidth]{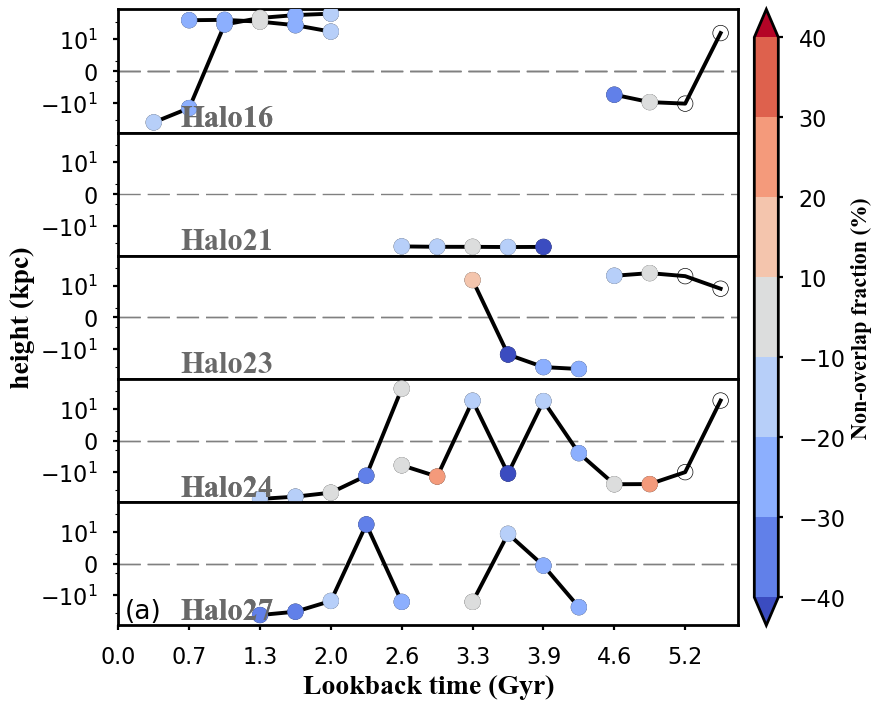}
    \includegraphics[width=0.4\linewidth]{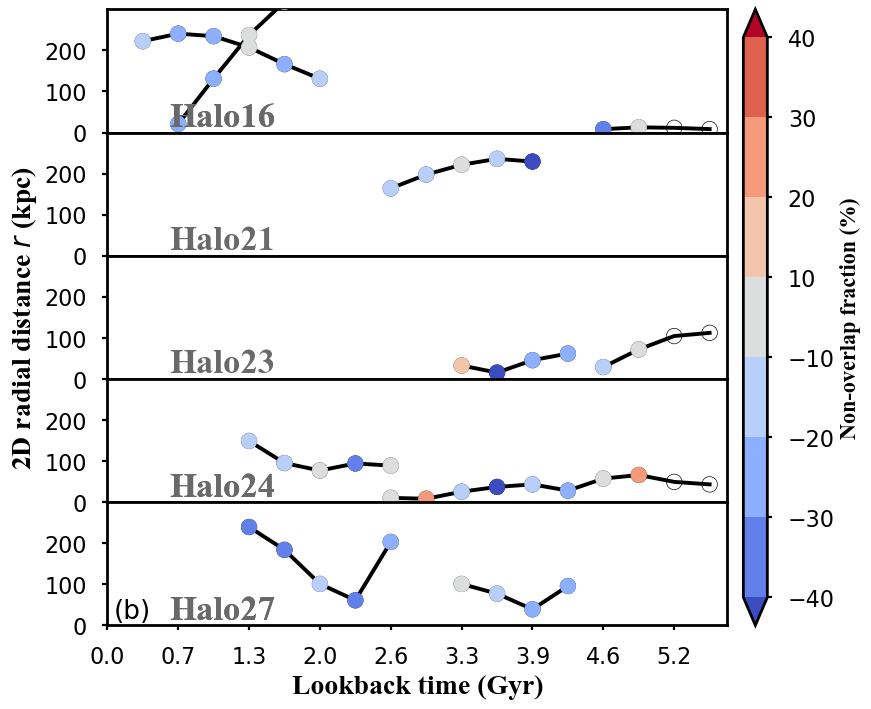}
    
    \includegraphics[width=0.4\linewidth]{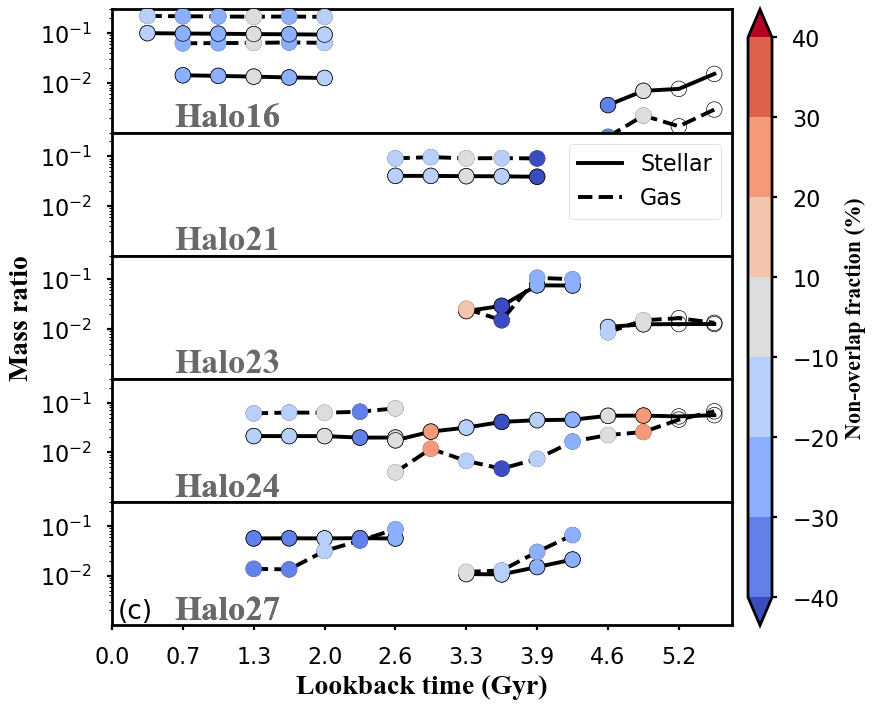}
    \includegraphics[width=0.4\linewidth]{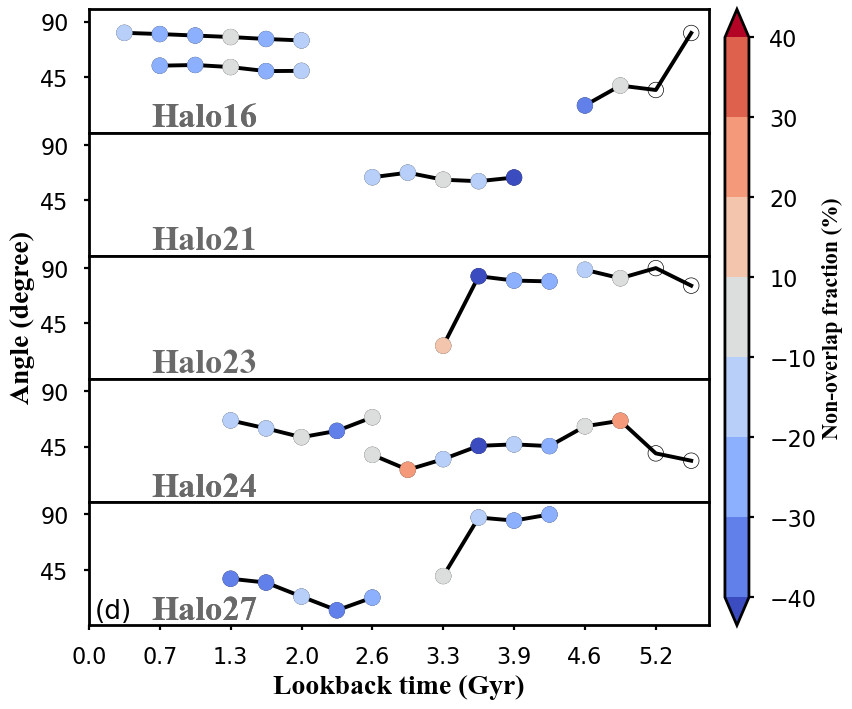}
    \caption{Height (panel a) and 2D radial distance (panel b) of satellites to their main haloes are presented. Panel c shows the gas (dashed) and stellar (solid) mass ratio. \revi{Panel d shows the acute angles between the spin axes of the main and satellite galaxies (0$^\circ$ corresponds to spins aligned along the same axis, while 90$^\circ$ is perpendicular to the disk).}
    Only satellites from snapshots corresponding to \ctwo\ and \cthree\ are included, with data points colour-coded by \fno\ as derived in Sec.~\ref{sec:fno_dmu}.
    We extensively present the orbital history and mass ratios beyond 5Gyr (hollow points) if a \ctwo\ or \cthree\ event occurs within three snapshots (1Gyr).}
    \label{fig:merge_prop}
\end{figure*}

If a given snapshot is aligned with \ctwo\ or \cthree\ and this snapshot occurs within 300~Myr (one snapshot) of a merging event, we trace the orbital history of the merged satellite in the previous $\sim$1.2~Gyr (four snapshots).
Similarly, if \ctwo\ or \cthree\ occurs during a satellite fly-by of the main halo, we track the satellite's orbit to two snapshots ($\sim$ 600~Myr) before and after \ctwo\ and \cthree\ appears. 
Fig.~\ref{fig:merge_prop}a summarizes the evolution of the satellites' height ($z$-component) relative to the main halo in the past 5~Gyr. 
The dashed lines represent the plane of the main halo's disc, i.e., height $h_\mathrm{z}= 0$. 
Fig.~\ref{fig:merge_prop}b presents the projected 2D radial distance ($r$) between the satellites and the main halo centre at each snapshot, traced back to 5~Gyr ago.
These two panels provide comprehensive satellite orbital history data.
When a satellite's orbit crosses the dashed lines in the top panel and $r$ is relatively small ($\lesssim$ 100~kpc for a conservative evaluation), the satellite has passed through the main halo's disc.
Fig.~\ref{fig:merge_prop}c summarises the gas (dashed) and stellar (solid) mass ratio between the satellites and their main haloes over cosmic time.
Fig.~\ref{fig:merge_prop}d investigates whether the angle of an interaction impact the age distributions.

We will discuss the presence of \ctwo\ in each halo in detail below.
The findings summarise that when an azimuthal age variation is absent, we observe a gas loss in companion galaxies or the presence of fly-by satellites with gas masses of $\sim$10\% of the main halo's mass.
Near the \ctwo\ snapshots, some satellites punch through the main halo's disc, while others do not. Therefore, our results do not suggest a direct correlation between orbital history and the disappearance of the age gradient. 
Instead, our data indicates that gas-rich interactions are the primary mechanism responsible for the similar age distribution on each side of the spiral arms.
We also find no clear preference for either \revi{parallel} or perpendicular interactions in \ctwo.
\revi{Most of \ctwo\ are observed to associate with gas stripping from satellites. It is possible that stripped gas may form new stars and contribute to the disruption of the azimuthal age gradient. In general, both gas stripping and tidal interactions likely play a role in shaping the age distribution across spiral arms.}

\subsubsection{Case~\textsc{ii} in Halo~16}
At a lookback time of $\sim$ 5~Gyr, a satellite with a stellar mass ratio of $\sim$1:100 and a gas mass ratio of $\sim1:150$ (orange line in the left panel of Fig.~\ref{fig:result_16}) crossed the main Halo~16's disc plane at small $r$ ($\lesssim$ 20~kpc) and merged. 
This satellite shows a continuous decrease in gas mass before merging into the main halo (Fig.~\ref{fig:result_16} and Fig.~\ref{fig:merge_prop}c).
The process of gas stripping may explain the absence of an azimuthal age distribution just before the merger event.
Later, \ctwo\ is observed again in Halo~16 at a lookback time of $\sim 1.3$Gyr, with no recent satellite crossings through the main halo’s disc.
Near this snapshot, two satellites are orbiting around Halo~16, one of which (blue line in the left panel of Fig.~\ref{fig:result_16}) has a stellar (gas) mass ratio of $\sim$ 1:10 (1:9).
The comparable gravitational potential in the satellite may give rise to the absence of azimuthal age distribution in the main halo.

\subsubsection{Case~\textsc{ii} in Halo~21}
In Halo~21, at a lookback time of 3.3 Gyr, a satellite (blue line in the left panel of Fig.~\ref{fig:result_21}) flies by and passes through the main halo's disc just before the \ctwo\ snapshot. 
The satellite has a high gas (stellar) mass ratio of $\sim$ 1:9 (1:10), with evidence of gas stripping between 3.3 and 4 Gyr ago. 
This interaction may have disturbed the density wave, leading to the absence of an age gradient across the spiral arms.

\subsubsection{Case~\textsc{ii} in Halo~23}
We find Halo~23 showing \ctwo\ at a lookback time of $\sim$ 4.9~Gyr. 
Near this \ctwo\ snapshot, there is a satellite with a stellar and gas mass ratio of $\sim$1:90 (green line in Fig.~\ref{fig:result_23}) orbiting above the mid-plane of the disc and eventually merging into Halo~23.
Similar to Halo~16, we observe a significant gas loss in the satellite before the merging event.
The absence of azimuthal variation in the stellar age can be attributed to the gas stripping process.

\subsubsection{Case~\textsc{ii} in Halo~24}
Halo~24 has encountered most satellites over the past 5~Gyr, with merger trees indicating that four satellites with stellar mass ratios \textgreater 1:100 have entered \Rvir. 
We identify three snapshots showing \ctwo\ at lookback times of around 2 Gyr, 2.6 Gyr and 4.6 Gyr.
A fly-by satellite (orange line in the left panel of Fig.~\ref{fig:result_24}) crossed the main halo's disc 2.5~Gyr ago, and 500 Myr later, an absence of an age gradient across the spiral arms in the main halo. 
This satellite has a gas-mass ratio of $\sim$1:45 and a stellar mass ratio of $\sim$1:90, compared to Halo~24.
Another satellite (dark red line in the left panel of Fig.~\ref{fig:result_24}) approached and orbited Halo~24 starting 6 Gyr ago, merging around 2.6 Gyr ago. 
This satellite crossed the main halo's disc multiple times before the merger and is accompanied by significant gas loss lasting for over 2 Gyr during the interaction.
The complex orbital history, involving multiple interactions and the final merger, likely contributed to the fluctuating \fno\ and \dmu\ in Halo~24 between 2.6 and 5~Gyr ago.

\subsubsection{Case~\textsc{ii} in Halo~27}
Halo~27 exhibits a younger leading edge in all snapshots during the past 5 Gyr, with a single exception at a lookback time of 3.3~Gyr, coinciding when a satellite galaxy merged (green line in the left panel of Fig.~\ref{fig:result_27}).
The satellite crosses the main halo’s disc and undergoes a significant decrease in gas mass right before the merger, with a gas (stellar) mass ratio of $\sim$ 1:45 (1:90). 
This satellite's transverse orbit likely facilitates gas stripping, transferring material to the main halo and subsequently erasing the younger leading edge signature.

\subsection{Case \textsc{iii}: Differences in the Broadness of the Age Distributions}\label{sec:case3}
In Halo~23, Halo~24, and Halo~27, we observe snapshots with noticeable discrepancies in the broadness of the age distributions between the leading and trailing edges, despite comparable mean ages (\cthree, bottom panel of Fig.~\ref{fig:age_dist}). 
Notably, one side of the spiral arms displays a broader age distribution (with a larger $\sigma$).
This suggests that while the mean age offset across the spiral arms is minimised, age asymmetry between the two sides remains. 
In this section, we discuss the orbital history and satellite properties of galaxy interactions near \cthree\ snapshots.

In Halo~23, we observe \cthree\ at a lookback time of 3.3~Gyr, coinciding with the merger of a satellite (blue line in the left panel of Fig.~\ref{fig:result_23}) with a stellar mass ratio of $\sim$ 1:30. 
This satellite crossed the main halo's disc just before the onset of \cthree\ (Fig.~\ref{fig:merge_prop}), with significant gas loss occurring around 600~Myr before the merger.
Similarly, 2~Gyr ago in Halo~27, a satellite (blue line in the left panel of Fig.~\ref{fig:result_27}) experienced gas stripping while crossing the disc, potentially linked to the disappearance of age gradients (small \dmu) across spiral arms. 
In Halo~24, at a lookback time of 5.3~Gyr, a satellite (dark red line in Fig.~\ref{fig:result_24}) approached the main halo, passing through the disc plane, and started showing a significant and continuous gas loss.
$\sim600$Myr (two snapshots) later, we observe very little age gradient (small \dmu) across the spiral arms, while age asymmetry persists as age distributions differ on either side of the spiral arms (large \fno).
This satellite merged into Halo~24 at the lookback time of 2.6~Gyr, after a few interactions.

All three \cthree\ snapshots occur under a common scenario: gas-rich merging or fly-by satellites traversing the main halo's disc.
These results indicate that this specific type of galaxy interaction disrupts the age distribution, erasing out the signal of the younger leading edge.
The variation in the broadness of the age distributions corresponds to a larger $\sigma$ on one side of the spiral arms, which means a more diverse stellar population on that side (leading edge for Halo~23 and 27 and trailing edge for Halo~24). 

We note, however, that \cthree\ is not observed after every gas-rich interaction with a traversing orbital history. 
This includes Halo16 at 5 Gyr ago, Halo21 at 3.3 Gyr ago, and Halo~27 at 2 Gyr ago. 
In these cases, the age distributions on both sides of the spiral arms overlap greatly, showing negligible differences in both mean age and broadness.

\subsection{Comparing spiral galaxies in observations and the Auriga simulation}\label{sec:obs_sim}
Comparison between simulations and observational data is crucial for testing theoretical frameworks and allowing us to identify the inside physical mechanisms needed in simulations. 
\red{Several previous studies \citep[e.g.,][]{Bottrell_2017,Rodriguez-Gomez_2019, Tang_2021} have carried out quantitative comparisons by generating mock observations that incorporate realistic observational effects.
Some pipelines exist, such as Simspin \citep{Harborne_2020}, although these pipelines require observational data from the telescope for calibration -- potentially a full research project on its own.
This work aims to present the intrinsic response of spiral discs and stellar age patterns to environmental companions across cosmic time, without introducing observational effects.
Within the scope of this paper, we provide context for the age distributions identified in Auriga and the observed trends reported in the literature.}

Observations have found azimuthal variations in stellar age in several nearby spiral galaxies including M~74 \citep{Sanchez-Gil_2011}, NGC~1566 \citep[][]{Shabani_2018}, NGC~628, NGC~3726, NGC~6946 \citep{Sakhibov_2021} and M~101 \citep{Garner_2024}.
These observations are consistent with the prediction of density wave theory, showing azimuthal propagation in stellar age.
Some observational studies \citep{Yu_2018, Abdeen_2022} also measure the pitch angle of spiral arms with multi-wavelength data, providing another vantage point as density wave theory predicts a looser spiral arm in the older stellar population (traced by redder wavelengths).
Indeed, some spiral galaxies in the local Universe \citep{Yu_2018, Abdeen_2022} and at higher redshift \citep[up to 8~Gyr ago;][]{Martinez-Garcia_2023} do show colour jump across the spiral arms.
To examine these phenomena in simulations, mock observations across multiple wavelengths \citep[e.g., TNG;][]{Lan_2024} or mapping 2D distributions of multi-phase ISM structures, such as CO and H$\alpha$ \citep[e.g., FIRE-2;][]{Hopkins_2018}, could provide valuable insights for future work.

The gas-phase ISM properties provide additional insight for examining spiral arm theories.
\citet{Ho_2017} presents a chemical model in a spiral galaxy driven by density waves, which leads to a lower metallicity -- 12+log(O/H) -- in the leading edge. 
Observational evidence for such metallicity offsets have been reported in NGC~1365 \citep{Ho_2017}, NGC~1087, NGC~1672 \citep{Kreckel_2019} and NGC~1566 \citep{Chen_2024b}.
Other gas-phase properties, including electron temperature \citep{Ho_2019} and star formation rate (SFR) \citep{Chen_2024}, also exhibit azimuthal offsets.
These findings indicate azimuthal star formation propagation across spiral arms, consistent with density wave theory. 
In the Auriga simulation, we observe a younger stellar population along the leading edge in most snapshots, even recovering a younger leading edge after a merger or fly-by event. 
The age azimuthal variations align with some observed spiral galaxies, consistent with the prediction of density wave theory.

In contrast to density wave theory expectations, some spiral galaxies are observed to have older stellar populations in the leading edge, such as M~51 \citep{Sanchez-Gil_2011} \footnote{We note that \citet{Sanchez-Gil_2011} observed azimuthal age variation in M~51, whereas \citet{Shabani_2018} reported little to no variation. The discrepancy may result from the difference in spatial resolution, as the \citet{Sanchez-Gil_2011} analysis is done on a pixel level whereas the \citet{Shabani_2018} study is done using the age of unresolved star clusters.} and NGC~2442 \citep{Chen_2024b}. 
Both M~51 and NGC~2442 are grand-design spiral galaxies, interacting with their companion (M~51b and NGC~2443, respectively). 
In this work, Halo~24 displays an older leading edge at a lookback time of 3~Gyr, $\sim$300~Myr before a satellite merges in. 
This snapshot presents grand-design spiral structures and prominent tidal features (Fig.~\ref{Afig:halo24}), resembling the characteristics seen in M~51 and NGC~2442.
This suggests that the merging event is likely a primary factor contributing to the age azimuthal variation observed in M~51 and NGC~2442.

Some spiral galaxies in the local Universe do not exhibit age gradients across their spiral arms.
The interacting systems M~51b \citep{Sanchez-Gil_2011} are reported to have no age pattern associated with spiral arms.
\citet{Choi_2015} find no age azimuthal variation in M~81, which experiences hydrogen gas stripping toward M~82 and NGC~3077.
\citet{Wezgowiec_2022} find no age azimuthal variation in NGC~628, with hot magnetic gas suggesting a possible tidal interaction.
We find little age azimuthal variation, i.e., \ctwo, in the snapshots undergoing gas stripping or gas-rich interactions (Sec.~\ref{sec:case2}).
Our findings in Auriga are consistent with these observations.
More importantly, this work highlights the potential role of gas-rich mini- and minor mergers in erasing detectable age gradients. 
\citet{Karademir_2019} discuss that mini-mergers frequently occur without disrupting the galactic disc. 
However, the detection of mini-mergers remains challenging due to the low surface brightness of the satellites. 
Our study provides a novel indirect diagnostic for identifying mini-mergers based on azimuthal age variations.

In a comparison of the gas-phase metallicity in NGC~2835, NGC~2997 and M~83, \citet{Chen_2024b} find negligible offsets in the median $\Delta$(O/H) (radial gradient subtracted metallicity) while the Kolmogorov–Smirnov test and Anderson-Darling tests suggest $\Delta$(O/H) on each side of the spiral arms support the distributions arising from different parent distributions.
NGC~2835 presents a flattening radial profile \citep{Chen_2023}, indicating accreted gas in the outskirts \citep[which is also seen in simulations;][]{Garcia_2023}.
Tidal features are observed in the H~\textsc{i} in both NGC~2997 and NGC~5236.
Similarly, in the Auriga simulation, we identify three \cthree\ snapshots where the leading and trailing edges have comparable average ages, but their broadness differs.
Our analysis of the merger tree (Sec.~\ref{sec:merger}) and orbital history (Sec.~\ref{sec:case3}) suggests that a gas-rich satellite passes through the main halo disc before \cthree\ snapshots.
The azimuthal distribution, similar to \cthree, along with the asymmetric gas distribution, suggests that NGC2835, NGC2997, and NGC~5236 may have undergone gas-rich interactions involving a crossing of the galaxy disc plane.
However, it is important to note that \citet{Chen_2024b} derives gas-phase metallicities from emission lines on a spaxel-by-spaxel basis. 
Potential offsets between leading and trailing edges may be obscured by significant scatters in gas-phase metallicities. 
Additionally, we emphasise that the analysis presented here focuses on stellar variations, whereas some prior observational studies have primarily examined variations in the gas phase.
Nevertheless, while this study focuses on star particles rather than gas cells, we expect both simulations and observations to capture the same underlying physical mechanisms.

Previous studies in both observations and simulations \citep{Neumann_2024, Marques_2025, Debattista_2025} suggest that bars can influence gas flows and stellar migration, potentially giving rise to azimuthal variations in stellar age. 
In Auriga simulations (Appendix~\ref{sec:all_images}), Halos~16, 23, and 27 show a bar over the past 5~Gyr with little evolution in size. 
Halo~21 develops a bar around 1~Gyr ago during interactions with two nearby satellites, while Halo~24 remains unbarred over the entire 5~Gyr period. 
Among all snapshots showing younger leading edges, 66\% are barred and 34\% do not; conversely, in all snapshots where bars are present, 83\% exhibit a younger leading edge. 
These results align with previous studies, indicating that the presence of a bar, together with spiral arms, can lead to azimuthal variations in stellar age.

\section{Conclusions}\label{sec:conclu}
We analyse five spiral galaxies from the Auriga L3 simulation dataset to investigate the long-term evolution of stellar age azimuthal distributions over the past 5~Gyr, focusing on environmental effects from mergers and fly-bys. 
To identify spaxels on spiral arms in each snapshot, we develop an automated ``ridgeline walking'' algorithm and divide the disc region into leading and trailing edges, following \citet{Chen_2024}. 

To measure azimuthal variations relative to the spiral arms, we compare the age distributions of young stars on each side of the spiral arms using two metrics: \fno\ (non-overlap fraction) and \dmu\ (mean age difference). These metrics quantify the differences in age distribution between the leading and trailing edges, providing insights into the impact of environmental interactions on spiral arm evolution.

We track fluctuations in both \fno\ and \dmu\ over the past 5Gyr.
Our results indicate that spiral galaxies in the Auriga simulations generally have younger leading edges compared to trailing edges, supporting density wave theory as the primary mechanism driving distribution in isolated spiral galaxies.
However, gas-rich interactions (mini-mergers or fly-bys) can erase azimuthal age offsets. 
We identify three snapshots where the mean age difference between the two sides of the spiral arms is negligible, yet significant differences appear in the age distribution tails.
All of these snapshots coincide with gas-rich interactions involving satellites crossing the main halo’s disc plane.

Comparing our results with observational data from previous studies, we find a consistent trend where galaxy interactions erase azimuthal variations. While $\sim$70\% snapshots in our study a younger leading edge, only $\lesssim$30\% of nearby galaxies exhibit azimuthal variations in metallicity and/or age gradient across spiral arms.

For future study, simulations with multi-phase ISM (e.g., FIRE; \citeauthor{Wetzel_2016} \citeyear{Wetzel_2016}) enable a more direct and detailed comparison with observations.
Simulations with a high temporal resolution (e.g., NIHAO; \citeauthor{Buck_2020} \citeyear{Buck_2020}, \citeauthor{Buder_2024} \citeyear{Buder_2024}) can also provide more constraints on the recovery timescale of age patterns.

\section*{Acknowledgements}
\revi{We thank the anonymous referee for their insightful suggestions.  }
QHC sincerely acknowledges Caroline Foster for her valuable advice and support in this work.
\red{The authors acknowledge Research Computing (\href{https://rc.virginia.edu}{https://rc.virginia.edu}) at The University of Virginia for providing computational resources and technical support that have contributed to the results reported within this publication.}

Parts of this work are supported by the Australian Research Council Centre of Excellence for All Sky Astrophysics in 3 Dimensions (ASTRO 3D), through project number CE170100013.
KG is supported by the Australian Research Council through the Discovery Early Career Researcher Award (DECRA) Fellowship (project number DE220100766) funded by the Australian Government. 
LCK acknowledges support by the DFG project nr. 516355818.
RSR and LCK acknowledge support from the ASTRO 3D visitor programme.
SB acknowledges support by DECRA funding through DE240100150.
\red{AMG and PT acknowledge support from NSF-AST 2346977 and the NSF-Simons AI Institute for Cosmic Origins which is supported by the National Science Foundation under Cooperative Agreement 2421782 and the Simons Foundation award MPS-AI-00010515.}

This research has made use of NASA's Astrophysics Data System Bibliographic Services (ADS). 
This research made use of {\sc astropy},\footnote{\href{http://www.astropy.org}{http://www.astropy.org}} a community-developed core Python package for Astronomy \citep{astropy13, astropy18}. 
This research has made use of the NASA/IPAC Extragalactic Database (NED) which is operated by the Jet Propulsion Laboratory, California Institute of Technology, under contract with NASA. 
Parts of the results in this work make use of the colormaps in the {\sc{cmasher}} package \citep{vanderVelden2020}.

\section*{Data Availability}
The data used in this article are available on the Auriga simulation website \href{https://wwwmpa.mpa-garching.mpg.de/auriga/index.html}{https://wwwmpa.mpa-garching.mpg.de/auriga/index.html}.
The products of this work are available upon request.
The ``ridgeline walking'' algorithm is available on GitHub \href{https://github.com/Qian-HuiChen/ridgeline_walking}{https://github.com/Qian-HuiChen/ridgeline\_walking}.



\bibliographystyle{mnras}
\bibliography{example} 



\appendix
\section{Azimuthal age patterns across different radii}\label{appendix:radial_bin}
\red{Under the framework of density wave theory, azimuthal age patterns are expected to reverse across the co-rotation radius -- from a younger leading edge inside co-rotation to an older leading edge beyond it. 
However, since the spiral structures in the Auriga simulations are not imposed as stationary density waves, it is challenging to determine a pattern speed and thus a co-rotation radius. 
In this section, we examine the azimuthal age variations at different radii without explicitly determining the co-rotation radius.}

\begin{figure*}
    \centering
    \includegraphics[width=0.33\linewidth]{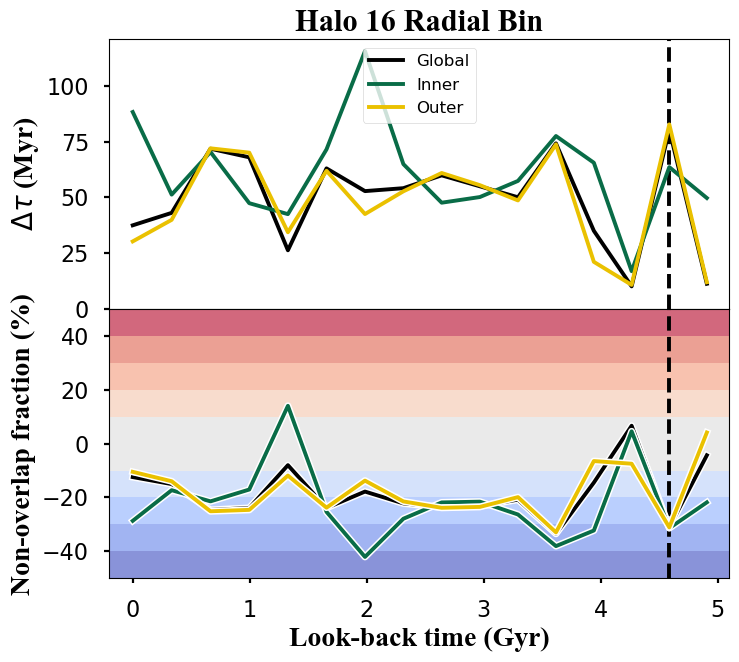}
    \includegraphics[width=0.33\linewidth]{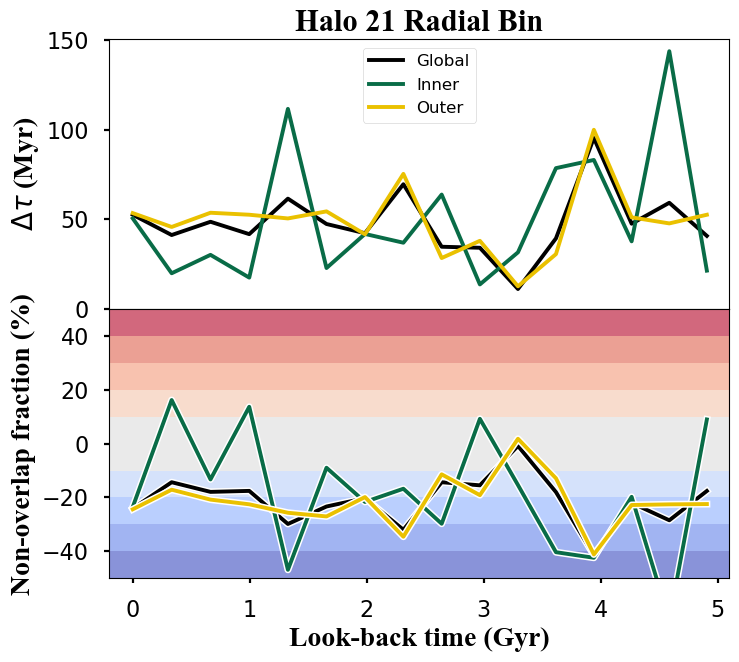}

    \includegraphics[width=0.33\linewidth]{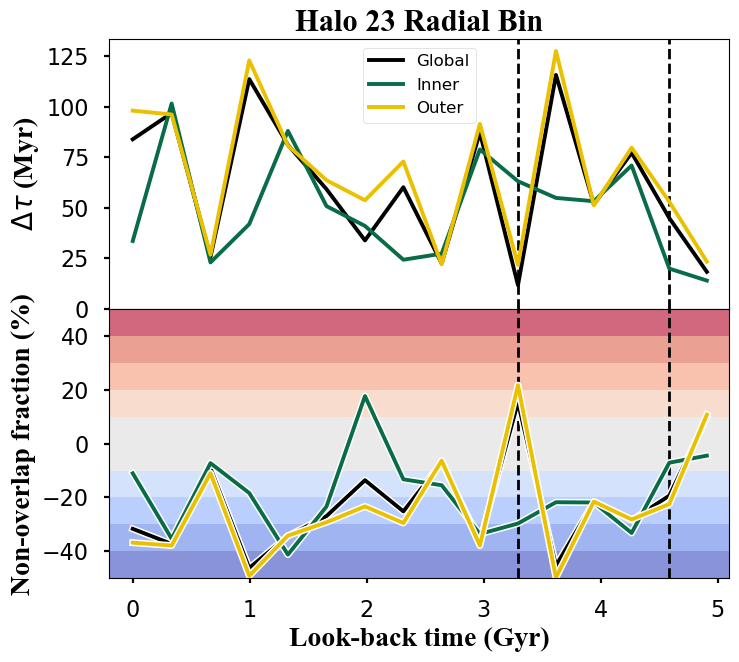}
    \includegraphics[width=0.33\linewidth]{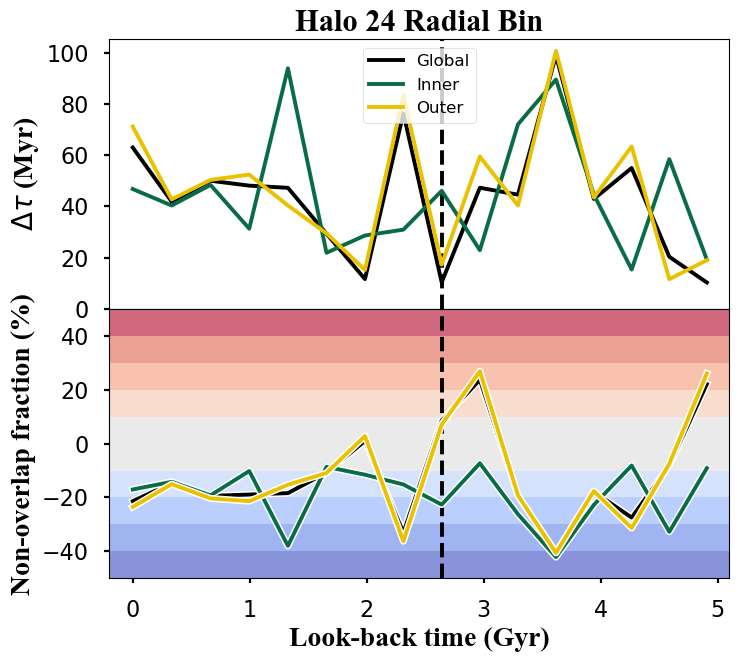}
    \includegraphics[width=0.33\linewidth]{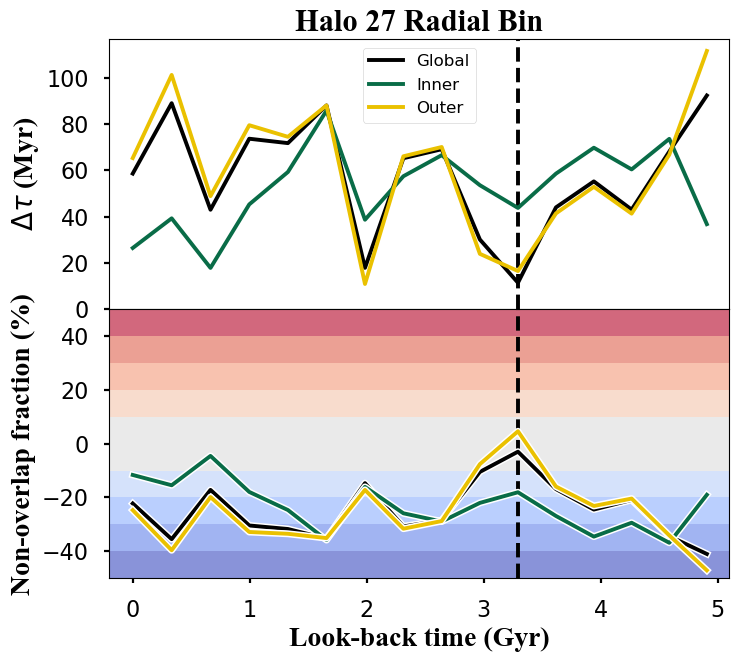}
    \caption{\red{Comparison of azimuthal age patterns at two radial bins -- $0-10$ kpc (inner) and $10-20$ kpc (outer). 
    The vertical dashed line marks a merger event and the black solid line shows the global trend discussed in the main text. 
    The inner and outer disc generally show qualitatively similar trends of age patterns. 
    We do not find a consistent trend for reversed age pattern across different radii -- a case where \fno\ falls in the red regime (older leading edge) in one radial bin while in the blue (younger leading edge) in the other radial bin.}}
    \label{fig:radial_bin}
\end{figure*}

\red{
The co-rotation radii measured in previous observational studies cover a broad range, with most of them lying between $5-15$ kpc \citep{Scarano_2013, Abdeen_2020}.
We divide the disc into two radial bins: $0-10$ kpc (inner) and $10-20$ kpc (outer). 
Figure~\ref{fig:radial_bin} compares the evolution of both parameters \dmu\ and \fno\ for each radial bin. 
For reference, the global trend discussed in the main text is shown as the black solid line.}

\red{In most snapshots, the inner and outer regions show generally consistent trends: both \fno\ fall in the blue regime and \dmu\ remains relatively high, indicating coherent age patterns across the disc, although we do observe slight quantitative differences between the two radial bins.}

\red{A few snapshots show different behaviours between the inner and outer disc. 
For example, the \fno\ fall into opposite regimes (one blue, one red) at 1~Gyr ago in Halo~21, 3.3~Gyr ago in Halo~23, and 3~Gyr ago in Halo~24. 
These discrepancies often coincide with fly-by or merger events, suggesting that such environmental interactions impact the inner and outer regions to different degrees. 
In particular, the outer disc is more strongly affected by the infall of accreted gas and stars, compared to the inner region.}

\red{We do not find strong evidence for reversed age patterns at different radii; in particular, there is no consistent trend of older leading edges in the outer disc. 
However, we note that in the unlikely case where the co-rotation radius lies beyond 20~kpc, such a reversal might not be captured within our current radial bins.}

\section{Comparison of L3 and L4 dataset}\label{appendix:L3vsL4}
The Auriga simulations release L3 and L4 datasets, which differ in baryon mass resolution \citep{Grand_2024}. 
In our main analysis, we utilise the L3 dataset, which has a higher baryon mass resolution of $6\times10^3$M$_\odot$, allowing for a more detailed representation of spiral structures.

\begin{figure}
    \centering
    \includegraphics[width=1\linewidth]{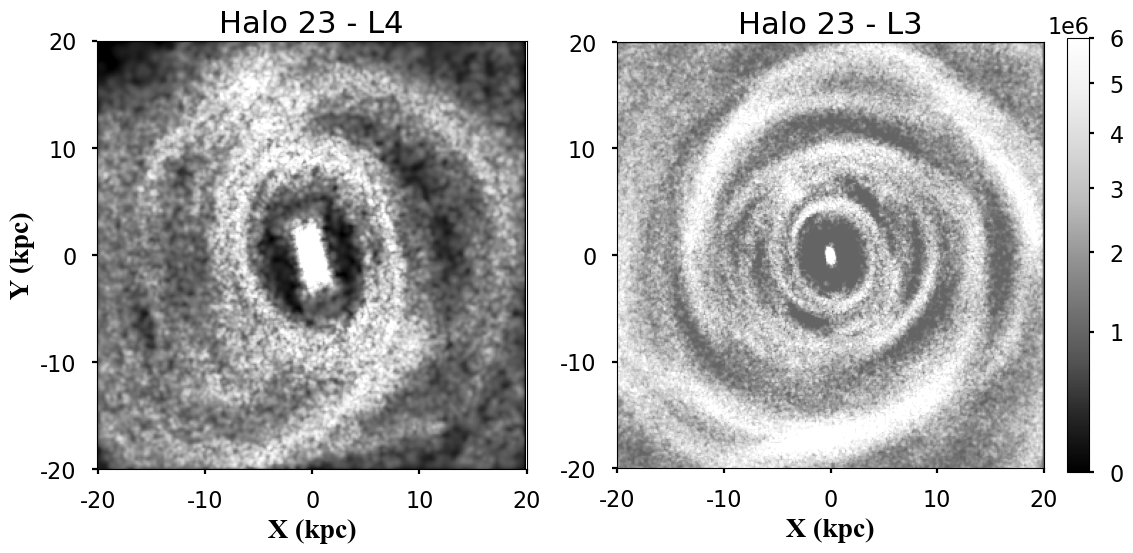}
    \caption{Young stars ($<2$~Gyr) mass map of Halo~23, simulated at L4 (left) and L3 (right) resolutions. 
    The visualisation process is described in detail in Sec.\ref{sec:auriga}. 
    The higher-resolution L3 dataset reveals more prominent and detailed spiral structures compared to L4.
    Notably, the lower-left spiral arm in the outer region, observed in L3, become absent in L4.}
    \label{fig:L3vsL4}
\end{figure}

The formation and evolution of spiral galaxies in simulations are resolution-dependent, with different resolutions yielding distinct spiral structures.
In the lower-resolution L4 simulation (Fig.~\ref{fig:L3vsL4}), spiral arms appear smoother and less prominent compared to the inter-arm regions, and tightly wound inner spirals transition into a ring-like structure.
As a result, the identification of spiral arms and the measurement of age gradients are inherently influenced by the baryon mass resolution.
While L4 provides a larger sample of halos, this study focuses on the higher-resolution L3 simulations to minimise resolution effects and ensure more robust structural analysis.

\section{Halo~6 exhibit tightly wound and non-prominent spiral arms}\label{appendix:halo6}
\begin{figure*}
    \centering
    \includegraphics[width=0.3\linewidth]{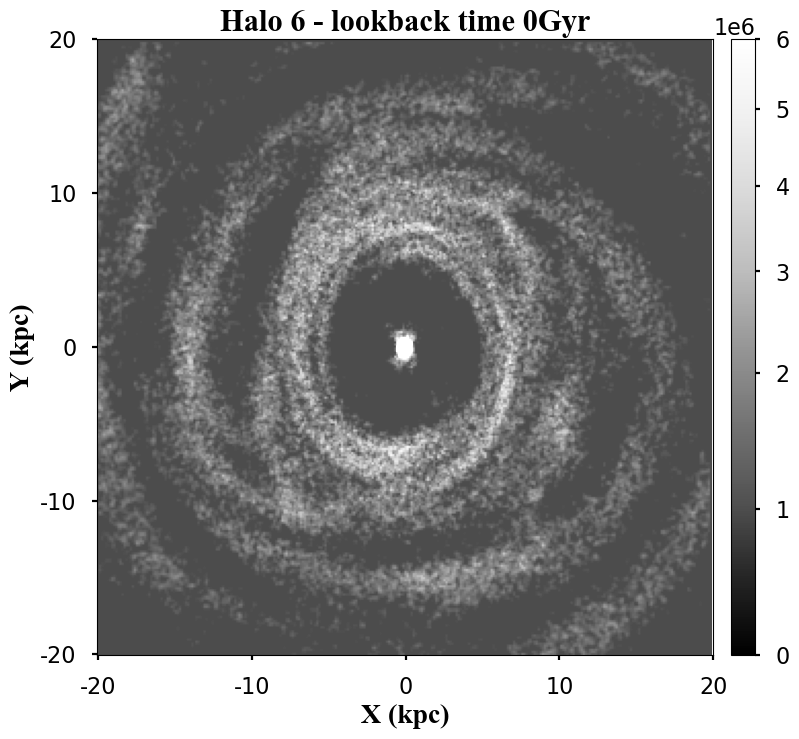}
    \includegraphics[width=0.3\linewidth]{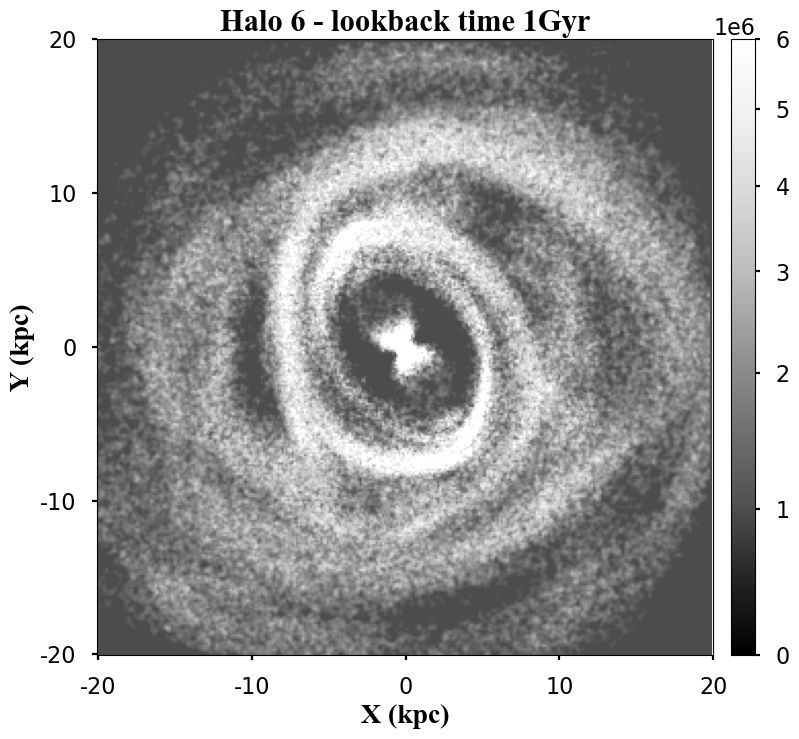}
    \includegraphics[width=0.31\linewidth]{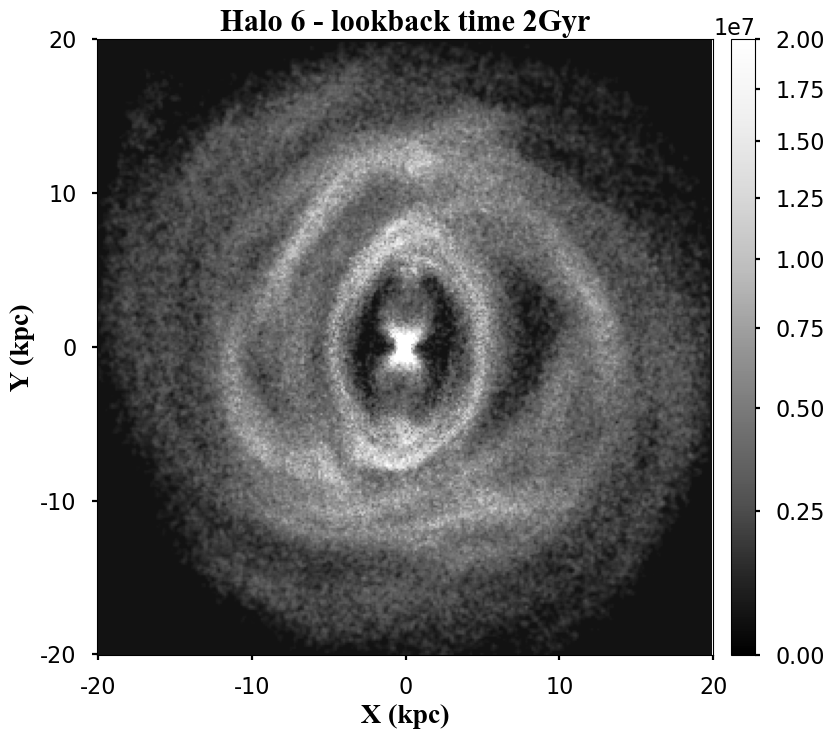}
    \caption{Young star ($<2$~Gyr) map of Halo 6 at the lookback time of 0~Gyr (left), 1~Gyr (middle) and 2~Gyr (right). }
    \label{Afig:halo6}
\end{figure*}

Fig.~\ref{Afig:halo6} presents young star maps of Halo~6 at lookback times of 0, 1, and 2 Gyr. 
Halo~6 features spiral arms with very small pitch angles, approaching a ring-like structure.
We find a gas-rich satellite (gas mass ratio of 1:5) within \Rvir\ of Halo~6 over the past 5~Gyr, which could attribute to the appearance of the ring structure \citep{Mazzei_2014, Mapelli_2015}.
These tightly wounded spiral arms make it challenging to distinguish the disc's leading and trailing edges. 
The spiral arms are non-prominent, blending smoothly with other young stars in the disc, making them difficult to locate visually or with the "ridgeline walking" algorithm (Sec.~\ref{sec:arm_def}). 
To maintain accuracy in defining spiral arms and comparing stellar ages on both sides of the arms, we exclude this halo from our analysis.

\section{Young star maps of each snapshot}\label{sec:all_images}
In this appendix, we present the young star ($<2$~Gyr) mass maps of all five halos in our sample over the past 5~Gyr, using the visualisation method described in Sec.~\ref{sec:aurigaL3}.
Spiral arms identified by the ridgeline walking algorithm (see Sec.~\ref{sec:arm_def}) are marked in red.

\begin{figure*}
    \centering
    \includegraphics[width=1\linewidth]{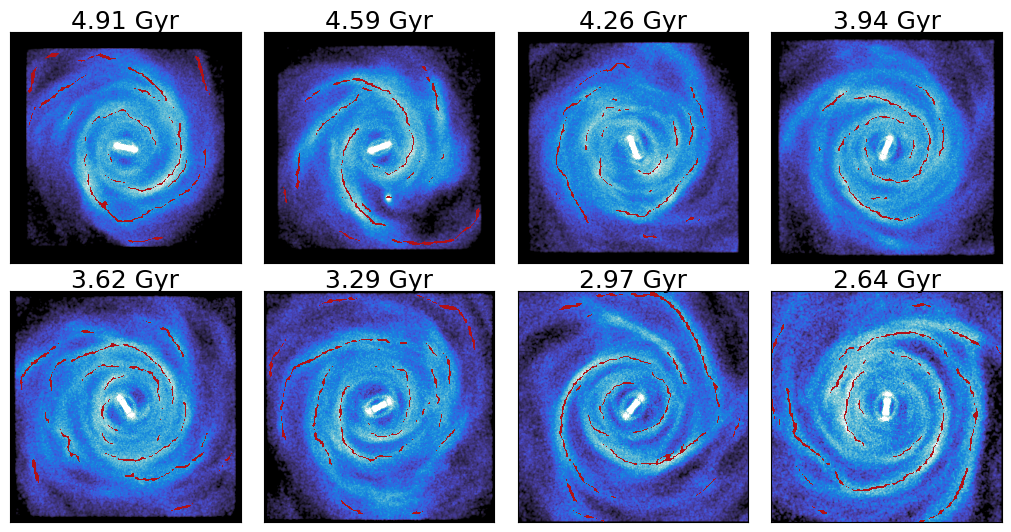}
    \includegraphics[width=1\linewidth]{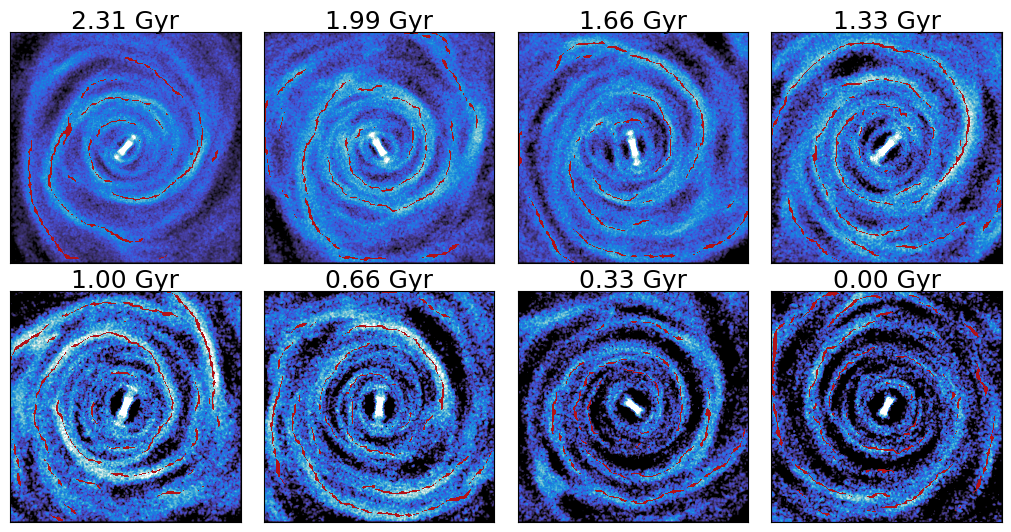}
    \caption{Young star (age $<2$~Gyr) mass map of Halo 16 over the past 5~Gyr, overlaid the definition of spiral arms found by ridgeline walking algorithm (Sec.~\ref{sec:arm_def}).}
    \label{Afig:halo16}
\end{figure*}

\begin{figure*}
    \centering
    \includegraphics[width=1\linewidth]{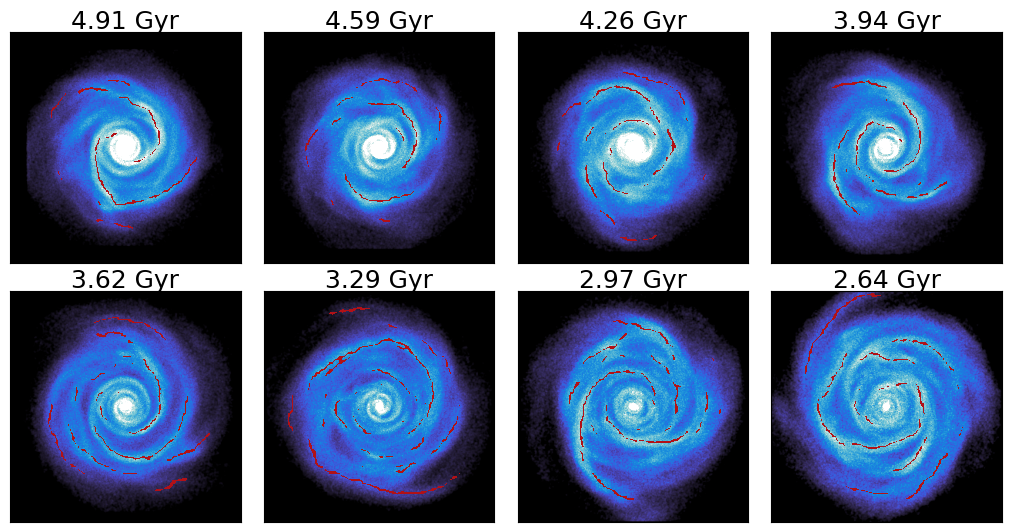}
    \includegraphics[width=1\linewidth]{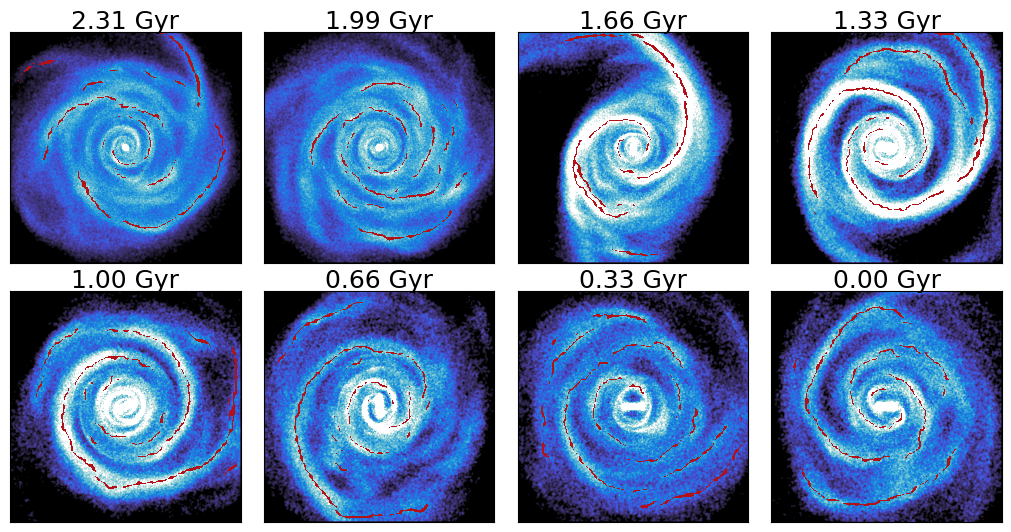}
    \caption{Similar to Fig.~\ref{Afig:halo16} but for Halo 21}
    \label{Afig:halo21}
\end{figure*}

\begin{figure*}
    \centering
    \includegraphics[width=1\linewidth]{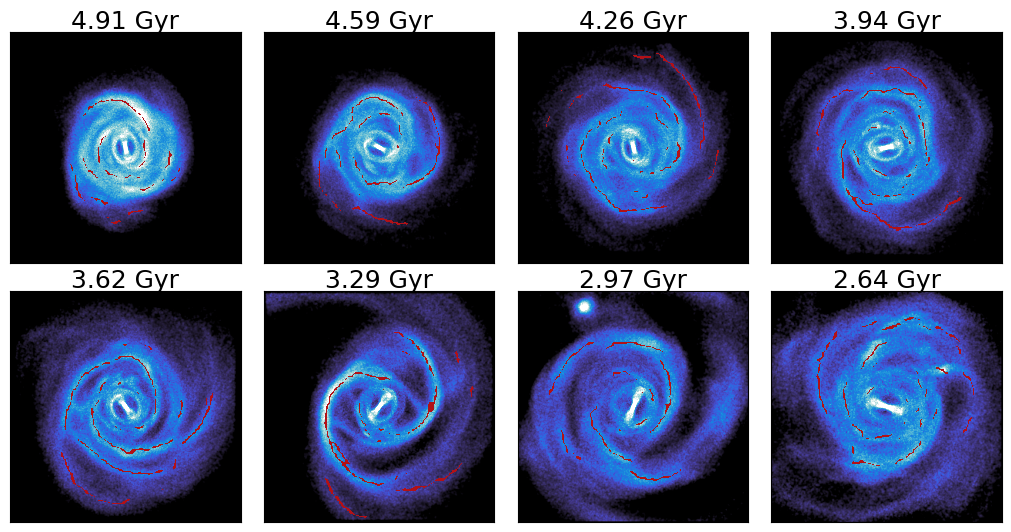}
    \includegraphics[width=1\linewidth]{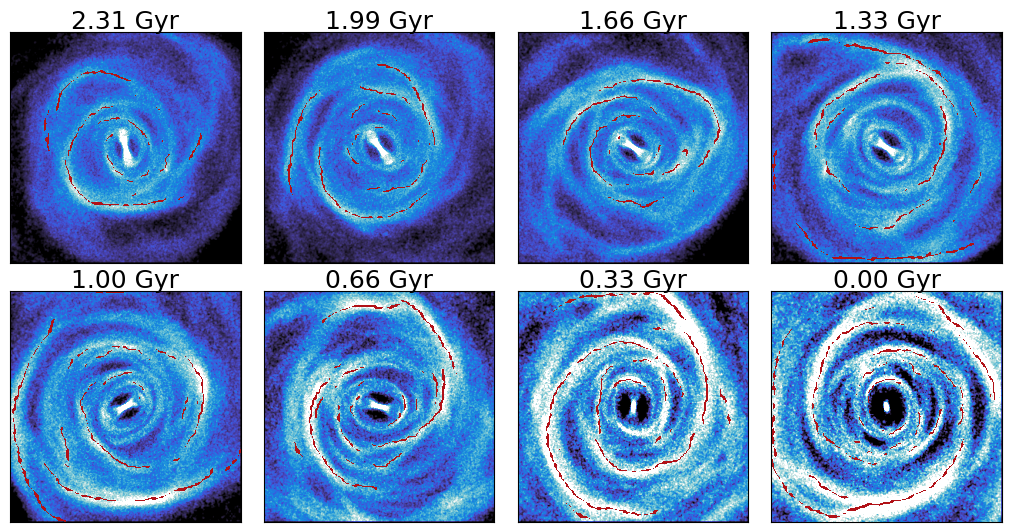}
    \caption{Similar to Fig.~\ref{Afig:halo16} but for Halo 23}
    \label{Afig:halo23}
\end{figure*}

\begin{figure*}
    \centering
    \includegraphics[width=1\linewidth]{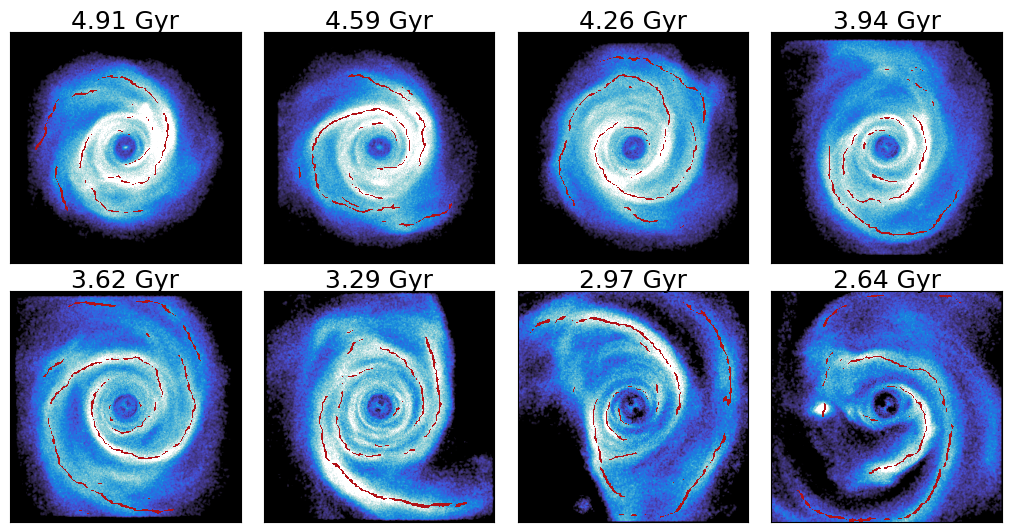}
    \includegraphics[width=1\linewidth]{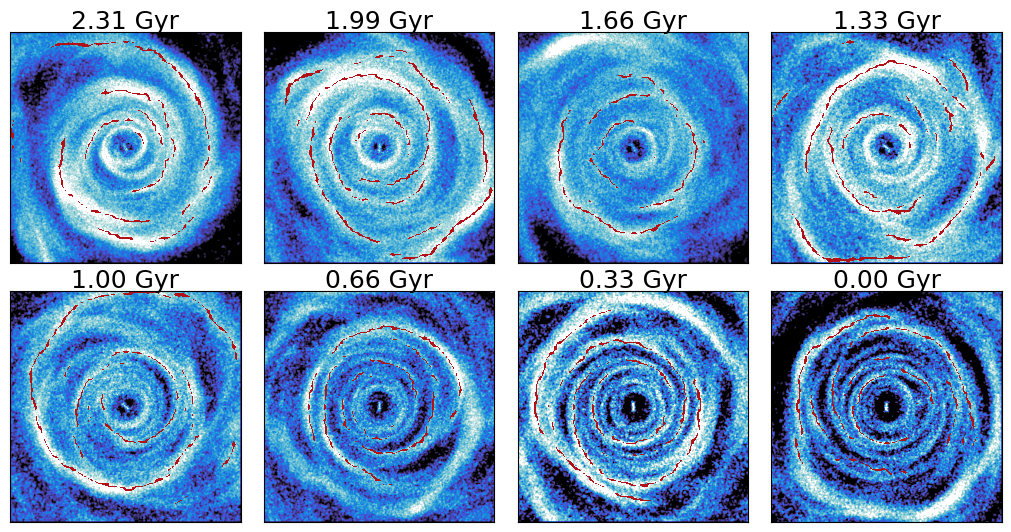}
    \caption{Similar to Fig.~\ref{Afig:halo16} but for Halo 24}
    \label{Afig:halo24}
\end{figure*}

\begin{figure*}
    \centering
    \includegraphics[width=1\linewidth]{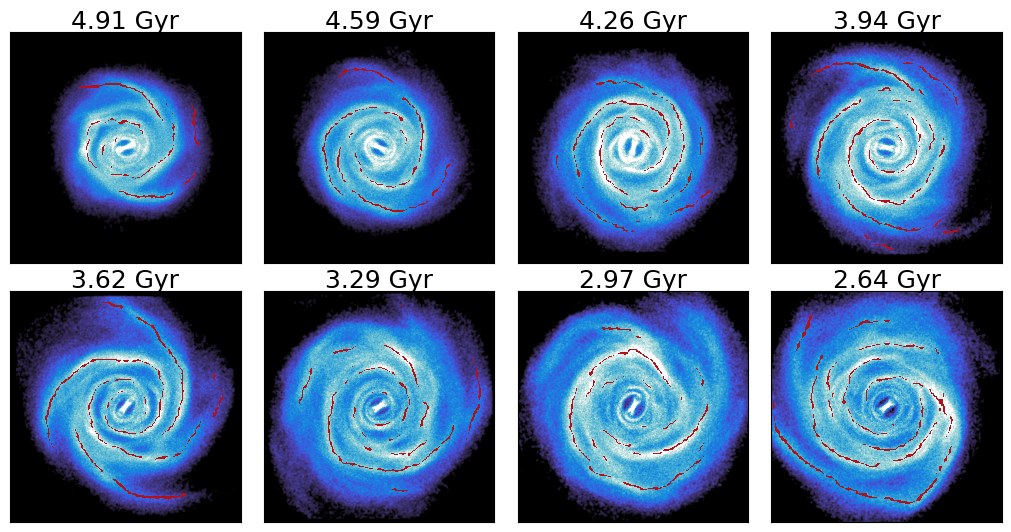}
    \includegraphics[width=1\linewidth]{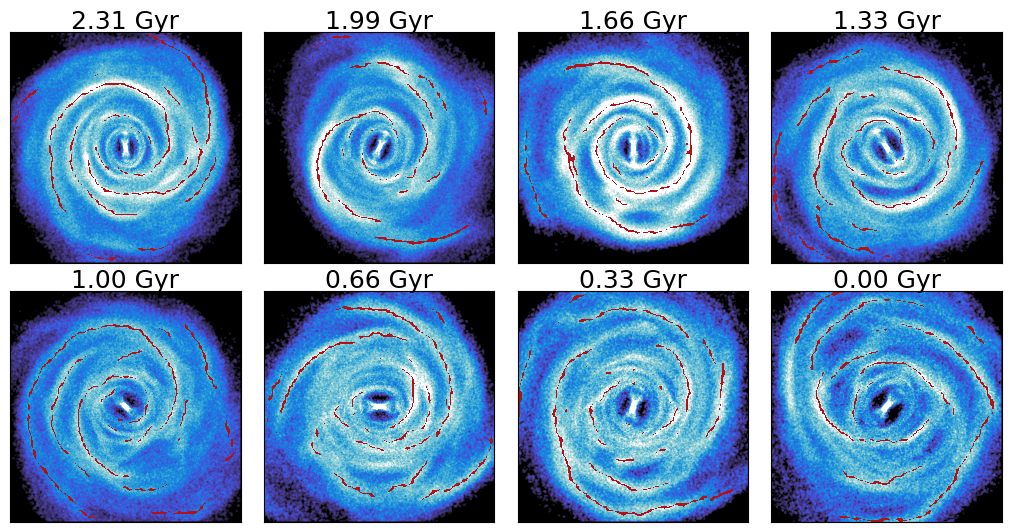}
    \caption{Similar to Fig.~\ref{Afig:halo16} but for Halo 27}
    \label{Afig:halo27}
\end{figure*}

\end{CJK} 
\label{lastpage}
\end{document}